\documentclass[iop]{emulateapj}
\usepackage{amsmath}

\def\t0{\theta_{\circ}}

\def\be{\begin{equation}}
\def\en{\end{equation}}

\def\msun{M_{\sun}}
\def\rsun{R_{\sun}}
\def\lsun{L_{\sun}}
\def\msunyr{M_{\sun} \, yr^{-1}}

\def\mdot{\dot{M}_{acc}}
\def\curf{{\cal F}}

\def\h2{H$_2$}
\renewcommand{\thefootnote}{\fnsymbol{footnote}}

\begin{document}

\title
{The evolution of accretion in young stellar objects:
Strong accretors at 3 -- 10 Myr*\footnotetext[1]{This paper includes data gathered with the 6.5 meter Magellan Telescopes located at Las Campanas Observatory, Chile.}}

\author{Laura Ingleby\altaffilmark{1}, Nuria Calvet\altaffilmark{1}, Jesus Hern{\'a}ndez\altaffilmark{2}, Lee Hartmann\altaffilmark{1}, Cesar Briceno\altaffilmark{2}, Jon Miller\altaffilmark{1}, Catherine Espaillat\altaffilmark{3}, Melissa McClure\altaffilmark{1}}

\altaffiltext{1}{Department of Astronomy, University of Michigan, 830 Dennison Building, 500 Church Street, Ann Arbor, MI 48109, USA; lingleby@umich.edu, ncalvet@umich.edu}
\altaffiltext{2}{Centro de Investigaciones de Astronom{\'i}a (CIDA), M{\'e}rida, 5101-A, Venezuela}
\altaffiltext{3}{Department of Astronomy, Boston University, 725 Commonwealth Avenue, Boston, MA 02215, USA; cce@bu.edu}

\begin{abstract}
While the rate of accretion onto T Tauri stars is predicted to decline with age, objects with strong accretion have been detected up to ages of 10 Myr.  We analyze a sample of these old accretors identified by having a significant $U$ band excess and infrared emission from a circumstellar disk.  Objects were selected from the $\sim$3 Myr $\sigma$ Ori, 4--6 Myr Orion OB1b and 7--10 Myr Orion OB1a star forming associations.  We use high resolution spectra from the Magellan Inamori Kyocera Echelle to estimate the veiling of absorption lines and calculate extinction for our T Tauri sample.  We also use observations, obtained with the Magellan Echellette and in a few cases the SWIFT Ultraviolet and Optical Telescope, to estimate the excess produced in the accretion shock, which is then fit with accretion shock models to estimate the accretion rate.  We find that even objects as old as 10 Myr may have high accretion rates, up to $\sim$10$^{-8}\;\msunyr$.  These objects cannot be explained by viscous evolution models, which would deplete the disk in shorter timescales, unless the initial disk mass is very high, a situation which is unstable.  We show that the infrared spectral energy distribution of one object, CVSO 206, does not reveal evidence of significant dust evolution, which would be expected during the 10 Myr lifetime.  We compare this object to predictions from photoevaporation and planet formation models and suggest that neither of these processes have had a strong impact on the disk of CVSO 206.
\end{abstract}

\keywords{Accretion, accretion disks, Stars: Circumstellar Matter, Stars: Pre Main Sequence}

\section{ Introduction}
\label{intro}
The timescales relevant for circumstellar disk evolution are key to consider for planetary formation scenarios.  For example, the surface density of gas in the disk determines the migratory path of young planets and establishes traps where planet migration may be halted \citep{kretke12}, so the gas dispersal time impacts the ability for a planet to survive.  Inner disk gas is difficult to observe directly because many lines are emitted in the far- ultraviolet \citep[FUV]{france12b,schindhelm12}.  Alternatively, the accretion of material from the disk onto the star provides a tracer of the amount of inner disk gas through the mass accretion rate, or $\mdot$.  In the current paradigm for mass accretion, the stellar magnetic field truncates the disk at a few stellar radii, where material falls onto the star along the magnetic field lines forming a shock at the stellar surface \citep{uchida84,hartmann94}.  \citet{calvet98} showed that the shock emits high energy photons which are re-processed in the accretion streams, producing an accretion spectrum peaked in the ultraviolet (UV).

T Tauri stars are identified as Classical T Tauri stars (CTTS) or Weak T Tauri stars (WTTS), accreting and non-accreting respectively, by observing H$\alpha$, thought to be formed in the accretion flows \citep{edwards94,muzerolle98,muzerolle01}.  Originally, CTTS and WTTS were distinguished by measuring the H$\alpha$ equivalent width \citep{herbig88}; however, the definition has evolved with the availability of high resolution spectra.  The equivalent width cutoff between CTTS and WTTS was shown to depend on the spectral type of the star \citep{white03,barrado03}.  Additionally, sources with wide wings or asymmetric line profiles were identified as accretors, whereas narrow ($<$200 km/s) symmetric lines are produced by chromospheric activity in non-accretors \citep{white01,natta04}.   The distinction between CTTS and WTTS typically distinguishes between Class II and Class III objects as well.  Class II and Class III refer to an infrared (IR) tracer of the properties of dust in the circumstellar disk, where Class II objects have full disks and Class III objects have little to no dust in the inner disk.  So far, only one known star has no indication of a disk in $\emph{Spitzer}$ 3.6--70 $\mu$m data but is accreting at a very low rate, MN Lup.  A soft X-ray excess and wide H$\alpha$ wings were detected as evidence for accretion.  This object is expected to be going through a very short lived phase in which the disk is optically thin and the last of the gas is accreting \citep{guenther13}.
        
Observations of young stars have revealed that accretion rates start high, up to $10^{-4}\;\msunyr$ during the early FU Ori outburst phase \citep{hartmann85,vorobyov06}, fall to approximately $10^{-8}\;\msunyr$ after the outbursts end \citep{calvet05} and drop to very low levels ($10^{-10}-10^{-11}\;\msunyr$) or stop completely by 10 Myr \citep{muzerolle00,espaillat08b,ingleby11b}.  Indeed the fraction of accreting objects in a given star forming region decreases with the age of the region \citep{fedele10}.  There are a number of ways in which magnetospheric accretion may be halted.  If the disk is truncated outside of the co-rotation radius, all the gas will disperse in a slow wind away from the star instead of accreting onto the star \citep{bouvier07b}.  Alternatively, the disk material may be depleted in both the inner and outer disk \citep{ingleby12}, stopping the inward flow of  material through the disk.  Processes for depleting the full disk include viscous evolution combined with photoevaporation by the central star \citep{clarke01,font04,alexander06,gorti09,owen10,owen11,owen12} or nearby hot stars \citep{adams04,anderson13} and clearing of disk regions by planet formation \citep{lubow06,zhu11}.  One or more of these processes may be ongoing in a circumstellar disk and each deplete the disk on different timescales \citep{rosotti13}.

Clearly, the disk and accretion properties are related and both are expected to evolve with age; the disk mass decreases (as well as the fraction of sources with disks) while the accretion rate drops \citep{hernandez07b,fedele10}.  Typically, older CTTS have low mass accretion rates; however very few CTTS remain at 10 Myr, so they have not been well studied.  In this paper we calculate accretion rates for 8 CTTS in the 3-- 10 Myr age range to illustrate that objects may continue accreting strongly out to 10 Myr, possibly posing an issue for disk evolution theory.  This span of ages covers the time when only half of the stars in a cluster retain disks ($\sim$3 Myr) to ages greater than the expected time for all sources to lose their disks, around 6 Myr \citep{hernandez08}.  These objects have $U$ band excesses over the expected $U$ band emission from a WTTS of the same spectral type, indicating ongoing accretion.  The $U$ band excess is a common way to measure $\mdot$, using calibrations between $L_U$ and the accretion luminosity $L_{acc}$ \citep{gullbring98}; however, it does not provide information on the shape of the excess in the ultraviolet (UV) or optical, which is important for accurate accretion rate estimates.  Obtaining a UV spectrum is time consuming and difficult to do for weak or distant objects, so here we use optical spectra containing the Balmer jump combined with accretion shock models to determine the shape of the accretion excess.  The Balmer jump provides crucial information about the slope of the excess extending into the UV.  In fact, the ratio of the continuum on each side of the Balmer jump has been shown to depend on the accretion rate \citep{herczeg08}.

In Section \ref{obs} we discuss our targets and the observations used in our analysis, including optical spectra from the Magellan Observatory, optical photometry from the Michigan-Dartmouth-MIT (MDM) Observatory and UV photometry from SWIFT when available.  Section \ref{analysis} outlines the steps involved in calculating accretion rates and Section \ref{results} presents the results from our analysis.  Finally, in Section \ref{discussion} we compare our results to theories of pre-main sequence evolution and highlight one of the oldest objects in our sample as a source with unexpected disk and accretion emission at an advanced age.

\section{Sample and Observations}
\label{obs}
\subsection{The Targets}
\label{targets}
The wide spread star forming complex in Orion offers an ideal opportunity to study evolving stars in diverse environments.  Outside of the Orion A and B clouds, containing the youngest stars like the dense Orion Nebular Cluster, lie older less extincted young stellar associations.  Here, we focus on a few sources in the $\sim$3 Myr $\sigma$ Ori, 4--6 Myr Orion OB1b and 7--10 Myr Orion OB1a regions (Table \ref{tabobs}).  These associations are interesting because they are older than the well studied, 1 Myr Taurus and 2 Myr Chamaeleon I star forming regions, but still contain CTTS.  Sources in $\sigma$ Ori (SO 540 and SO 1036) were selected from Class II sources in \citet{hernandez07b} and were expected to be accreting.  Objects in OB1a (CVSO 206 and OB1a 1630) and OB1b (CVSO 58, CVSO 90, CVSO 107 and CVSO 109) were identified as accreting in \citet{briceno05}.  Orion OB1b and OB1a targets were also classified as Class II objects in \citet{hernandez07a}.  Our OB1a objects reside in a distinct group within OB1a, 25 Ori \citep{briceno07}.  

\begin{figure}[htp]
\plotone{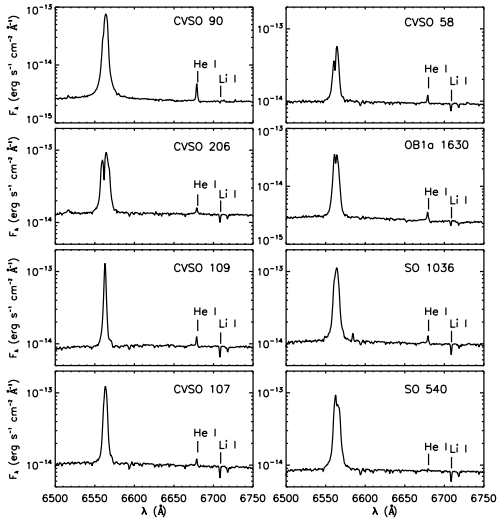}
\caption{Youth indicators in CTTS sample.  A portion of the MagE spectrum is shown for each source in the sample revealing the presence of H$\alpha$ and the 6707.8 {\AA} Li I absorption line.  The Li I line in CVSO 90 is weak due to veiling (see Section \ref{veiling}).  He I emission at 6678 {\AA} is an indicator of accretion detected in the MagE spectra.}
\label{halphli}
\end{figure}

The membership of the CIDA Variability Survey of Orion (CVSO) objects to the subassociations Orion OB1a or 1b was determined in \citet{briceno05} and \citep{briceno07}; they were initially selected as variable objects that fell above the zero age main sequence in a $V$ versus $V-I_c$ diagram
and were confirmed as members based on the depth of the Li I $\lambda$6707 line and/or other indicators of youth.  In Figure \ref{halphli} we show a portion of the MagE spectra of these stars including the Li I and H$\alpha$ lines, in agreement with those results.  Membership was further confirmed using radial velocities in \citep{briceno07}.  Object OB1a 1630 was selected as a member of 25 Ori from its location in the region
of Class II sources in the $\emph{Spitzer}$ Infrared Array Camera (IRAC) [4.5] - [5.8] versus [5.8] - [8.0] and [3.6] - [5.8] versus [4.5] - [8.0] diagrams  \citep{hernandez07a}.  Here, we confirm its membership from the presence and depth of Li $\lambda$6707 (Figure \ref{halphli}).  Membership of SO 540 and SO 1036 to $\sigma$ Ori was established by \citet{caballero08} and \citet{sacco08} based, again, on the strength of the Li I line, which we confirm with our spectra in Figure 1.  In addition, their colors in the  [4.5] - [5.8] versus [5.8] - [8.0] and [3.6] - [5.8] versus [4.5] - [8.0] diagrams showed them to be Class II  objects \citep{hernandez07b}.

In this paper, we consider that the age of the object is the same as the mean age of the population to which it belongs.  The $\sigma$ Ori cluster has been well studied in the literature and the range of ages estimated is 2--4 Myr \citep{zapatero02,oliveira02,sherry04}.  \citet{briceno05} estimated the ages of the subassociations OB1b and OB1a taking into account uncertainties in the distance and depth of each association and in the isochrones used, resulting in a range from 7.4 to 10 Myr for 25 Ori and 5.5 to 6 Myr for Orion OB1b.

The sample also includes 5 WTTS of varying spectral types used for comparison to the CTTS spectra, listed in Table \ref{tabwtts}.  WTTS were selected from $\sigma$ Ori, Orion OB1b and Orion OB1a when available and were confirmed to be WTTS based on high resolution H$\alpha$ spectra discussed in the next section.  CHXR 48 is a WTTS in the Chamaeleon I star forming region and was included due to the lack of a suitable late type WTTS in the regions of interest in this analysis.

\subsection{Spectral Observations with MIKE and MagE}
We observed 8 CTTS and 5 WTTS with the Magellan Echellette (MagE) and the Magellan Inamori Kyocera Echelle (MIKE) on the Magellan Clay telescope at Las Campanas Observatory (LCO) during November 2009.  The combination of instruments provided long wavelength coverage, with medium resolution spectra (R$\sim$4100) extending down to 3300 {\AA} with MagE and high resolution of fainter red absorption lines with MIKE (R$\sim$35,000).  We focus on data from the red arm of MIKE which covers 4900 to 9500 {\AA} and do not include the blue (3350--5000 {\AA}) because MIKE does not have the needed sensitivity in the blue.  MagE and MIKE data were reduced using the Image Reduction and Analysis Facility (IRAF) tasks CCDPROC, APFLATTEN, and DOECSLIT \citep{tody93}.  Due to the variable nature of T Tauri stars, we observed each source with MIKE and MagE as near to each other as possible, ideally observations were separated by 15 minutes to 3 hours.  On some occasions, due to unforeseen weather complications, sources were observed with MIKE and MagE on subsequent nights.  

\subsection{Photometry with MDM}
In order to flux calibrate the MagE spectra we coordinated the Magellan observing run with observations at the MDM observatory near Tucson, AZ.  Given variable weather conditions at the two sites, we were able to get photometric observations within one night of our spectral observations.  We obtained optical photometry with the Ohio State University Blue 4K CCD
(hereafter 4K imager) on the MDM 1.3 m McGraw-Hill telescope. The 4K imager has a 21\arcmin.3 square field
of view. We used 2 x 2 binning, which gives a plate scale of 0.\arcsec63.
The 4k imager has a known issue of crosstalk between the four CCD
segments. When a CCD pixel is saturated, it creates spurious point
sources in corresponding pixels on the other three segments.
The exposure sequence consisted of one set of short exposures
(30, 20, 20 and 15 seconds) and one set of long exposures
(300, 240, 180 and 180) in the $U$, $V$, $R$ and $I$ photometric bands.
Landolt standard fields were obtained each night for photometric
calibration in the Johnson photometric system.
Since the readout of the 4K imager uses four amplifiers, we have
four overscan regions in each CCD frame. First, we applied an overscan correction
using the IDL program proc4k written by Jason Eastman.
We then performed the basic reduction following the standard
procedure using IRAF. We obtained aperture photometry with
the IRAFphot package, using an aperture of 8", or $\sim$3.4 times the FWHM.
The rms departures of the standard stars from the calibration equations
are $\sim$0.03 mag for $V$, $R$ and $I$ bands, and $\sim$0.08 for $U-V$ color.  Observed magnitudes are provided in Table \ref{tabflux}.  CTTS are known to vary on timescales from hours to years.  While high cadence light curves
are not available for the objects in this sample, \citet{briceno07} found that typically CTTS
vary in $V$ by $\sim0.6$ magnitudes.  Due to time differences in our Magellan and MDM observations
this introduces an uncertainty in the flux of $\sim50\%$.

\subsection{Producing the MagE spectrum}
MagE is an echellette and therefore the orders must be pieced together in order to produce a continuous spectrum over all wavelengths.  To correct for the sensitivity function across each order we also observed flux standards during our observing program.  Ideally, the flux standards were observed at the same airmass and near the same time as the targets.  These flux standards are part of the European Southern Observatory's Optical and UV Spectrophotometric Standard Stars sample, which provides flux calibrated spectra over the wavelength range we are interested in.  We compared each order of the observed flux standard to the spectrum in the catalog and calculated the correction between the observation and catalog spectrum.  This same correction was then applied to our T Tauri stars.  This step does not flux calibrate the T Tauri star spectra but does remove the sensitivity function of each order.  In order to flux calibrate our source spectra, we scaled the MagE spectrum to photometry obtained at MDM as close to our MagE observations as possible.

\subsection{SWIFT/ UVOT Photometry}
Accretion emission is best observed in the UV, where the shock excess peaks and the stellar photosphere is dim. We obtained UV observations of a few of the CTTS in this sample with the Ultraviolet and Optical Telescope (UVOT) on the SWIFT Gamma Ray Burst Explorer in SWIFT GI program  \#6090725 (PI N. Calvet).  We chose fields of view with SWIFT that covered as many young stars as possible, while avoiding bright stars which were not safe for the UVOT detector.  The UVOT camera is ideal for accretion rate analysis because it has three filters in the NUV (UVW1, UVM2 and UVW2) as well as $U$, $B$ and $V$ filters and observations in all 6 filters could be obtained in approximately one hour.  The bandpass of each UV filter is 2200 to 4000 {\AA} for UVW1, 2000 to 2800 {\AA} for UVM2 and 1800 to 2600 {\AA} in UVW2.  Unfortunately the UVW1 and UVW2 filters have a red leak, where the tail of the transmission curves extends well into optical wavelengths.  As late type stars our objects are very red, especially for those with low $\mdot$, and therefore a significant fraction of the UVW1 and UVW2 fluxes comes from the star instead of the NUV excess produced by the shock.  Alternatively the UVM2 filter at 2221 {\AA} does not have a red leak and therefore the observed flux in that band is expected to be accurately probing the NUV excess.  Our SWIFT observations were not simultaneous with the optical data; however, with $U, B, V$ observations from SWIFT and contemporaneous $U$ and $V$ photometry from MDM we may determine whether there was significant variability between the optical and NUV sets of observations.

The SWIFT UVOT observations were analyzed using UVOTBADPIX, UVOTEXPMAP and UVOTDETECT.  UVOTBADPIX finds bad pixels in the sky images and creates a bad pixel map.  UVOTEXPMAP takes the bad pixel map and creates an exposure map in sky coordinates which gives areas in the image flagged as bad an exposure time of 0.  Finally, UVOTDETECT uses the exposure map made in the previous step to identify sources detected with a signal to noise ratio of 3.  UVOTDETECT also extracts the count rate for each detection from a region defined by an ellipse.  We compared the list of $V$ band detections to lists of known T Tauri stars \citep{briceno05,hernandez07a}.  Two sources, one in Orion OB1a and one in $\sigma$ Ori, were observed with all instruments in this analysis, including MIKE, MagE, the MDM 4k imager and SWIFT.  The remaining sources where not in the SWIFT fields of view due to proximity to bright stars.  A log of all observations may be found in Table \ref{tabobs} while UVOT magnitudes are in Table \ref{tabflux}.

\section{Analysis of Observations}
\label{analysis}

\subsection{Spectral Types}
\label{spt}
Although spectral types were previously determined for our sample \citep{briceno05,briceno07,hernandez07b}, here we re-determine them based on the MagE spectra.  We use the code SPTCLASS (SPecTral CLASSificator code) which classifies late K and M stars based primarily on TiO  molecular bands, with an uncertainty of $\pm$ 1 subclass \citep{hernandez04}.  The spectral types derived here (listed in Table \ref{tabprop}) agree with those from the literature, within the errors.  Similarly, we use SPTCLASS to confirm the spectral types of the WTTS used as templates in our analysis (see Table \ref{tabwtts}).  CTTS with large continuum excesses produced in the shock may be incorrectly classified.  The addition of the excess emission to the stellar photospheric spectrum produces shallower photospheric absorption lines \citep{hartigan89}, complicating spectral typing.  SPTCLASS may assign an earlier spectral type to heavily veiled objects, in particular CVSO 90, so the given spectral type may be considered an early-spectral type limit \citep{hsu12}.  \citet{manara13} showed that by not accounting for strong veiling, objects may be misclassified by as much as one spectral class.

\subsection{Calculating veiling in MIKE spectra}
\label{veiling}
\begin{figure}[htp]
\plotone{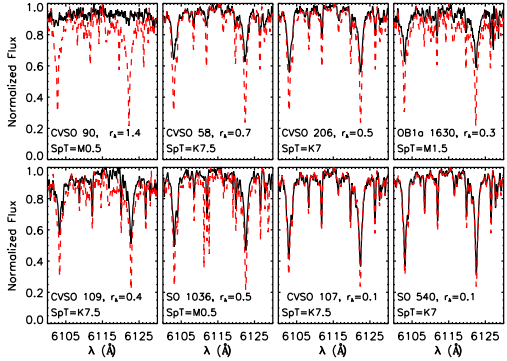}
\caption{Observed veiling in MIKE spectra between 6100 and 6130 {\AA}.  The black solid line in each panel is the CTTS and the red-dashed line is a WTTS template of the same spectral type.  We list the measured veiling in this region at the bottom.  Strong Ca I absorption lines are observed at 6102.7 and 6122.2 {\AA} \citep{merle11}.  Li I absorption is also detected against the Ca I line at 6103.5 {\AA} in most sources. }
\label{veilplot}
\end{figure}

It is important to determine the extent to which the absorption lines are veiled because it provides information of the relative contributions to the spectra from the emission of the accreting material and the underlying stellar photosphere.  Figure \ref{veilplot} shows examples of veiled absorption lines for the sources in our sample.  We measure veiling as the ratio of the continuum to the line depth \citep{hartigan89}.  Although in principle one could get an estimate of the excess from only one line, in practice many lines are used to minimize the uncertainties in veiling determinations.  Differential veiling over a short wavelength range and the possibility of emission cores imposed on photospheric absorption lines contribute to errors in our veiling estimates \citep{gahm08,dodin12}.  Therefore, we measure veiling across the entire available MIKE spectrum, following the methods of \citet{gullbring98}, by splitting the MIKE spectra into pieces of $\sim$15 {\AA}, avoiding any known emission lines produced by shock emission.  We use a WTTS of the same spectral type as a template with the continua of both the WTTS and CTTS normalized to unity.  Continuum emission is added to the spectrum of the WTTS in each 15 {\AA} interval in small increments between some minimum and maximum amount of veiling determined by eye and then renormalized by dividing by $1+r_{\lambda}$, where $r_{\lambda}$ is the veiling per wavelength.   A reduced chi squared test, $\chi^2_{red}$, is calculated for each spectrum of the WTTS plus the continuum compared to the CTTS and the fit with the lowest $\chi^2_{red}$ is assumed for that interval.

\begin{figure}[htp]
\plotone{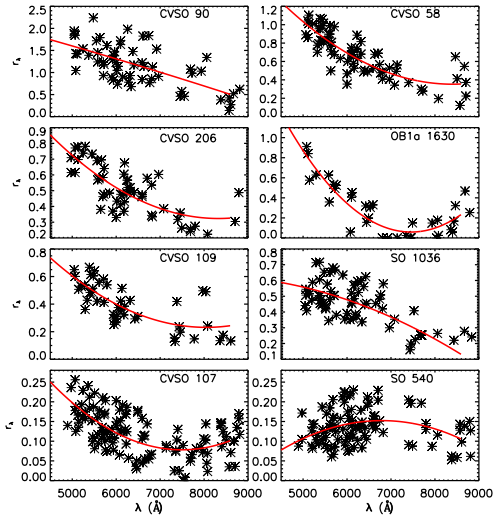}
\caption{Veiling versus wavelength measured in MIKE spectra.  For all stars except CVSO 109 and OB1a 1630 we calculated veiling using two different WTTS as templates.  We found poor fits with a second WTTS template for CVSO 109 and OB1a 1630 and therefore included only one template in the analysis of these objects.  The red line is a fit to the points, which we use to determine veiling at a given wavelength.}
\label{veilspect}
\end{figure}

This process is completed for each interval and the best fit is examined by eye and either accepted or rejected.  We are not using models, but real data as templates, so there are unforeseeable imperfections  in the templates.  For most sources, we used two different WTTS with spectral types equal to or within $\pm$0.5 subclasses to confirm that veilings were not dependent on the WTTS template chosen.  Finally a veiling ``spectrum" was made and a 3rd order polynomial was fit to determine the function of veiling versus wavelength (Figure \ref{veilspect}).  The line was first fit to all the data and then the standard deviation from that line was determined.  A new line was fit to the data within one standard deviation of the first fit in order to omit any spurious veiling measurements.   In most cases veiling decreases as wavelength increases, due to the shock emission peak in the UV.   From this fit, we determined the veiling ($r$) at $V$, $I$ and extend the function to estimate veiling at $J$.  Once veiling is known, we have an estimate of the excess emission through $r_{\lambda}=F_{\lambda,e}/F_{\lambda,p}$, where $e=$excess is the shock excess producing the veiling and $p=$photosphere is the flux of the underlying star.  We use veiling at $V$ and $I$ to estimate extinction and veiling at $J$ to determine the stellar luminosity in the next section.  Errors on $r_V$ and $r_I$ are listed in Table \ref{tabprop}.

\subsection{Extinction and Stellar Parameters}
While these objects reside in regions with low extinction, $A_V$'s calculated from optical colors are $>$0.  A significant problem with determining $A_V$ from optical colors is that the emission produced in the shock alters optical colors from the underlying photospheric colors.  Therefore, the commonly used method of comparing $V-I$ of the CTTS to $V-I$ for a standard star of the same spectral type has error for the CTTS.  We did use $V-I$ for the WTTS as there is no accretion excess to contaminate optical bands.  For CVSO 127 and SO 774, we obtained $V$ and $I$ photometry with MDM.  Photometry for 2MASS J05264681+0226039 and CVSO 173 came from the CIDA variability survey.  The $A_V$ for CHXR 48 is from \citet{luhman04}.

 For the CTTS, we used $V-I$, after correcting them for the amount of veiling measured in the MIKE spectra and $F_{\lambda,star}=F_{\lambda, obs}/(1+r_{\lambda})$, where $F_{\lambda, obs}$ is the observed flux.  By correcting $V$ and $I$ by the estimated veiling at each wavelength (listed in Table \ref{tabprop}), we have information regarding the photospheric emission below the shock excess and are able to discern how reddened the photospheric fluxes are.  Using corrected $V$ and $I$ and comparing to the standard colors from \citet{kenyon95} we estimate the extinction for our sample of T Tauri stars.  Errors in $A_V$ are dominated by spectral type uncertainty.    Recently, \citet{pecaut13} re-evaluated the colors of dwarf stars and they differ from the \citet{kenyon95} colors particularly for mid K to early M spectral types.  The new colors would decrease our estimated $A_V$'s by 0.2--0.4 magnitudes.  When calculating the accretion rates in Section \ref{results2}, the lower values of $A_V$ would decrease $\mdot$ by a factor of 1.8--2.4.  Here, we use the \citet{kenyon95} colors to calculate $A_V$ for better comparison of our accretion rates to values calculated previously but note the error introduced by $A_V$ estimates.  

To calculate stellar parameters we used the calibration between spectral type and effective temperature in \citet{kenyon95}.  We used 2 Micron All Sky Survey \citep[2MASS]{skrutskie06} de-reddened $J$ magnitudes, de-veiled using our estimated veiling from the MIKE data, and the bolometric correction per spectral type from \citet{kenyon95} to estimate bolometric luminosities.  The masses of the stars were estimated using the evolutionary tracks of \citet{siess00} and the stellar radii were calculated assuming distances of 440, 400 and 330 pc for $\sigma$ Ori, OB1b and OB1a, respectively.  All stellar parameters may be found in Table \ref{tabprop}.  The MagE spectra and MDM photometry were de-reddened using the calculated $A_V$ and the reddening law towards HD 29647 discussed in \citet{whittet04}.

\subsection{Determining the Excess using the Veiling}
\begin{figure}[htp]
\plotone{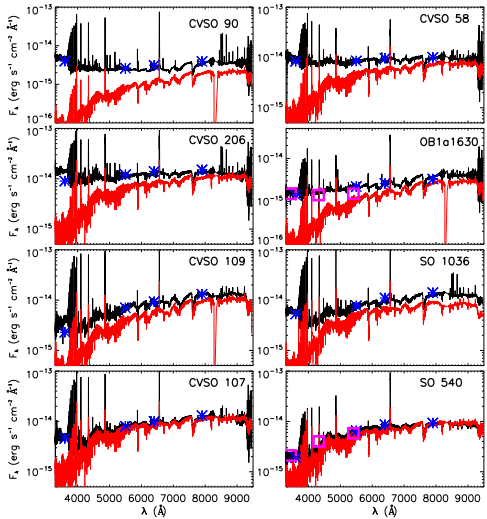}
\caption{MagE observations of CTTS sample.  The de-reddened CTTS spectra (black) are plotted along with WTTS templates (red).  MagE spectra of CTTS were scaled to the contemporaneous MDM photometry (blue asterisks) for flux calibration.  When available, we include SWIFT $U,V,B$ photometry (magenta squares).  Errors in the photometry are smaller than they symbol size.}
\label{cttswtts}
\end{figure}

To calculate the excess (or veiling) spectrum produced in the accretion shock, we scaled a WTTS of the same spectral type to the CTTS based on the veiling at $V$, calculated from the MIKE spectra.  Figure \ref{cttswtts} shows the MagE spectra of the CTTS sample and scaled WTTS templates.  We then subtracted the WTTS from the CTTS and located continuum regions in the spectrum of the shock excess.  We include the excess short of 3600 {\AA}, and between 3800 and 8000 {\AA} omitting wavelengths with balmer emission lines, Ca II h and k lines, and Na D lines.  The fits are weighted toward the blue end of the spectrum, where the shock excess is stronger.  These regions were used to fit the shock models to the excess spectrum (described below).  

\section{Calculating Accretion Rates}
\label{results}
\subsection{Description of Shock Model}
The disks of CTTS are inwardly truncated at the radius where the magnetosphere intercepts the disk.  Here, the strong magnetic field lines channel disk material onto the surface of the star.  The material reaches the star traveling at approximately free fall velocities, forming a shock which irradiates both the star below (post-shock and heated photosphere) and the material falling along the accretion column (pre-shock) with soft X-rays.  These regions reprocess the X-ray emission and the re-emitted emission peaks at UV or blue optical wavelengths \citep{calvet98}.

An in depth description of the accretion shock model may be found in \citet{calvet98}, where these models were first introduced.  \citet{calvet98} assumed that the shock emission was formed in a single column of accreting material, perpendicular to the stellar surface, characterized by a single energy flux ($\curf$) and filling factor ($f$) on the stellar surface.  Motivated by models of the magnetosphere geometry \citep{donati08,gregory11,gregory12}, \citet{ingleby13} showed that including multiple accretion columns, characterized by a range in energy flux and filling factor, were capable of fitting both blue, as well as red excesses observed in CTTS \citep{edwards06,fischer11}.  To do so, high $\curf$, low $f$ columns were assumed to co-exist with low $\curf$, high $f$ columns.  We again use the assumption that there are accretion columns with a range in properties when we fit accretion shock models to the current optical sample.  

\subsection{Results}
\label{results2}

\begin{figure}[htp]
\plotone{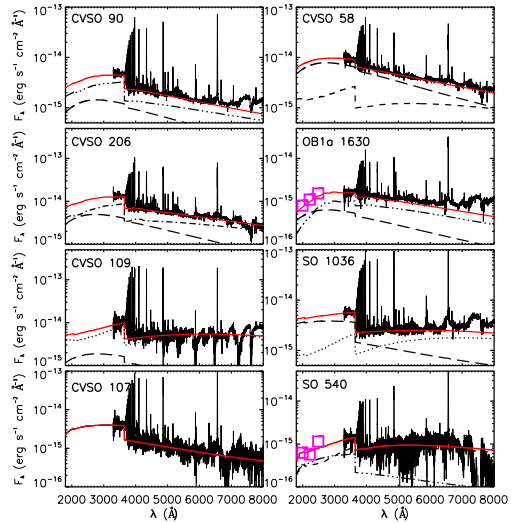}
\caption{Excess produced in the accretion shock.  MagE spectra of each CTTS is plotted with the WTTS template subtracted, leaving only emission produced by accretion.  The dashed lines represent different accretion shock models characterized by log $\curf$=10 (dotted), 10.5 (short dashed), 11 (dot-dashed), 11.5 (dash-triple dotted), and 12 (long dashed) erg s$^{-1}$ cm$^{-2}$.  The red solid line is a sum of all the accretion columns which contribute to the final shock model.  The magenta boxes on OB1a 1630 and SO 540 represent SWIFT photometry. }
\label{fitshock}
\end{figure}

\begin{figure}[htp]
\plotone{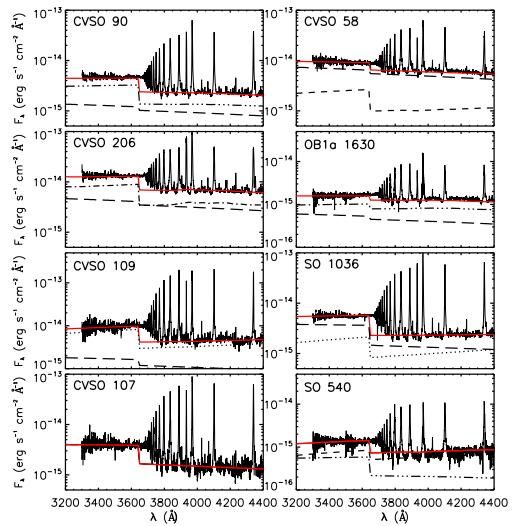}
\caption{Accretion shock fits to Balmer jump.  Spectra and lines are as defined in Figure \ref{fitshockzoom} but the Balmer Jump is zoomed in on to show the fits of the models in this region.  All sources, except OB1a 1630, have larger Balmer jumps than predicted by the standard model and additional optically thin pre-shock material is needed to reproduce the observations.}
\label{fitshockzoom}
\end{figure}

We fit the excess spectrum of each object using accretion shock models with log $\curf$=10, 10.5, 11, 11.5 and 12 
 erg s$^{-1}$ cm$^{-2}$.  For each $\curf$ model we varied the filling factor and added the contribution from each column.  We calculated the $\chi^2_{red}$ of each model fit to the CTTS excess spectrum to determine which combination of accretion columns gave the best fit to the MagE excess spectrum.  We found that the standard accretion columns could not accurately fit
most spectra in the region of the Balmer jump, around 3600 {\AA}, with
not enough excess produced at the bluest wavelengths.  Objects with
larger Balmer jumps can be explained by assuming lower electron
densities in single temperature slab models or a larger emitting area
for the low density preshock emission in the accretion shock models
\citep{herczeg08}.  We found that in all objects, with the exception
of OB1a 1630, larger preshocks were needed with up to 5$\times$ more
preshock emission necessary.  While it has not been shown where the additional 
low optical depth material giving rise to this excess emission resides it is 
likely that the preshock emission is underestimated here due to the assumed simple
geometry of the accretion columns.  From models fitting Zeeman
splitting observations, it is known that the magnetosphere is complex,
including higher order magnetic fields \citep{donati08,gregory11}.
Therefore, the footprints of these fields on the stellar surface are
likely complex and variable as well \citep{long11}.
 
 The total accretion rate for each object is the sum of the accretion rates for each individual column,
 \be
 \mdot=\frac{8\pi R_{\ast}^2}{v_s^2} \curf f
 \en
where $v_s$ is the velocity of the infalling material and $R_{\ast}$ is the radius of the star.  Our best fits to the shock excess are shown in Figure \ref{fitshock} with a close view of the Balmer jump region shown in Figure \ref{fitshockzoom}.  When available we also plot the SWIFT W2, M2 and W1 photometry, keeping in mind that the red leak in the W1 and W2 filters causes the flux to be over-estimated and that the M2 filter is the most reliable for this reason.  While we did not explicitly fit the SWIFT photometry, it agrees well with the fit to the optical spectra even though they are not simultaneous.  The red leak is not apparent in the higher accretor, OB1a 1630; however, the contamination in the short and long wavelength filters is observed in SO 540, where the accretion luminosity is low.  Properties of the accretion shock model for each source are provided in Table \ref{tabfill}.  Including errors in mass, radius and $A_V$, uncertainties in $\mdot$ are approximately a factor of 2.  Here we are not attempting to fit the emission lines produced by accretion, only the excess continuum.  We note that several of the CTTS have a remaining red excess, CVSO 90, OB1a 1630 and SO 1036, in particular.  These excess peaks are due to imperfect matches between the CTTS and WTTS, a result of slight differences in effective temperature or surface gravity.  Mismatches between the CTTS and WTTS are more easily observable at red wavelengths, where the shock contribution is low.  Another possibility is that the excess emission from dust in the disk is contributing a red excess; however, in most cases the flux from the disk  is significantly lower than the stellar flux near 8000 {\AA} \citep{mcclure13a}.

\section{Discussion}
\label{discussion}
\subsection{Evolution of the accretion rate}
\label{evolution}

\begin{figure}[htp]
\plotone{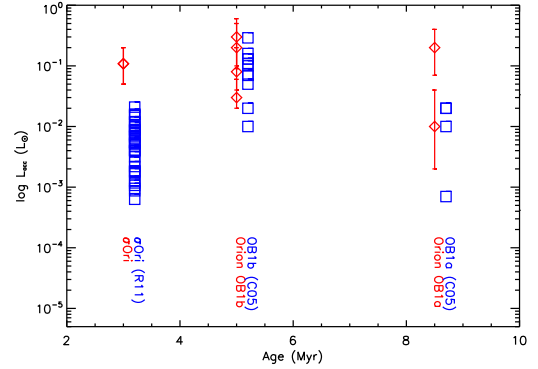}
\caption{Accretion luminosity versus age.  Red diamonds are accretion luminosities calculated in this paper.  Blue squares  show accretion luminosities calculated for $\sigma$ Ori \citep[R11]{rigliaco11}, Orion OB1b and Orion OB1a \citep[C05]{calvet05}.  One Orion OB1a transitional disk plotted, CVSO 224, has a low accretion luminosity of 0.0007 $\lsun$ \citep{espaillat08b}.}
\label{laccage}
\end{figure}

The objects presented in this paper are interesting because at
 their age they would be expected to be weakly accreting; however,
 they have accretion rates similar to objects in the 1 -- 2 Myr Taurus
 and Chamaeleon I star forming regions.  Figure \ref{laccage} compares the accretion luminosities, $L_{acc}$, found in this analysis to previously estimated $L_{acc}$ values for $\sigma$ Ori \citep{rigliaco11}, Orion OB1b and Orion OB1a \citep{calvet05,espaillat08b} for objects with spectral types between K6 and M3.  We select a small mass range for comparison due to the dependence of $\mdot$ on stellar mass \citep{muzerolle05}. The literature values were obtained using the $U$ band excess, which while not as accurate as multi-wavelength observations, provides an estimate of the total excess.  The $L_{acc}$  for Orion OB1b objects differ from the values calculated by \citet{calvet05}, primarily for CVSO 90 and CVSO 58.  This is due to the inclusion of veiling in our estimates for $A_V$ and when determining the contribution from the star to the total flux.  The highest accretors in $\sigma$ Ori were not included in the sample of \citet{rigliaco11} due to saturation in the $U$ band photometry.  There may also be a bias against detecting the lowest accretors in Orion OB1b because it has the highest average $A_V$ of the three regions.  Clearly there is a large spread in accretion properties for sources of a given age, with a range of 2--3 orders of magnitude in $L_{acc}$.

\begin{figure}[htp]
\plotone{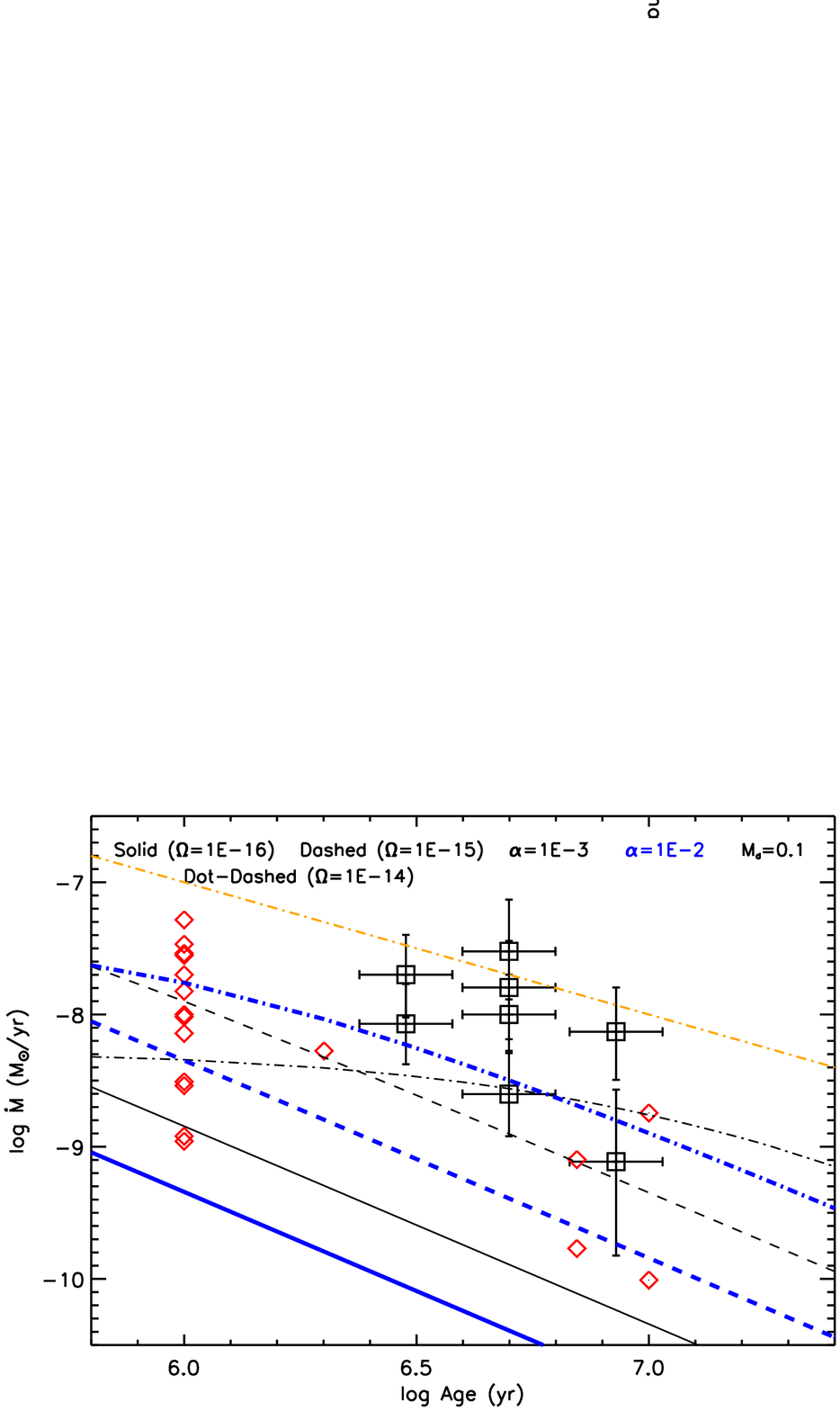}
\caption{Accretion rates compared to viscous evolution models.  The accretion rates measured in this analysis (purple diamonds) are combined with the CTTS sample of \citet{ingleby13}, shown as red diamonds.  The predicted evolution of $\mdot$ from Equation \ref{h98} is shown for $M_d=0.1\;\msun$, $M_{\ast}=0.5\;\msun$, $R_{\ast}=2\;\rsun$ and $T_{100}$=10 K, constant in all models.  The parameters $\alpha$ and $\Omega$ are varied, with $\alpha=10^{-3}$ in thin black, $\alpha=10^{-2}$ in thick blue, $\Omega=10^{-16}$ as the solid lines, $\Omega=10^{-15}$ as the dashed lines and $\Omega=10^{-14}$ rad s$^{-1}$ as the dotted lines.   None of the viscous evolution scenarios can reproduce high accretion at 10 Myr.  The orange dash-dotted line represents the $\mdot$ needed for a source to have accreted 0.1 $\msun$ during its lifetime.}
\label{mdotage}
\end{figure}

To investigate the range of star and disk properties responsible for
creating the large distribution in $L_{acc}$ and to eliminate differences
introduced by the method of obtaining the accretion properties, we
compare values of $\mdot$ for the older objects presented here to those
calculated for other star forming regions
using the same methods employed in this paper \citep{ingleby13}, again using objects with similar masses to the Orion samples.  In particular, \citet{ingleby13} used the same accretion shock model to determine
accretion rates for a significant sample of objects in the 1 Myr
Taurus region, as well as one object in the Chamaeleon I star forming region (at $\sim$ 2 Myr) and a few older objects in the
8 Myr $\eta$ Cha and 10 Myr TW Hydra Association (TWA). The samples of accretors in $\eta$ Cha and TWA are almost complete as these are small associations with only a
few known CTTS and no strong accretors.  Figure \ref{mdotage} shows $\mdot$ from this work and \citet{ingleby13} versus the age of the association or group.

Figure \ref{mdotage} also shows the predicted change of $\mdot$
onto the star due to viscous evolution of the circumstellar disk
(calculated at the disk truncation radius, $R=5\;\rsun$.), following \citet{hartmann09} and \citet{hartmann98};
\be
\begin{split}
\label{h98}
\mdot&=6\times10^{-7}\;\frac{e^{-R/R_1t_d}}{t_d^{3/2}}\left(1-\frac{2R}{R_1t_d}\right)\left(\frac{M_d(0)}{0.1\;\msun}\right)\\
&  \times\left(\frac{R_1}{10\;\rm{AU}}\right)^{-1}\left(\frac{\alpha}{10^{-2}}\right)\left(\frac{M_{\ast}}{0.5\; \msun}\right)^{-1/2}\\
&\times \left(\frac{T_{100}}{10\;\rm{K}}\right) \msunyr,
\end{split}
 \en
where $M_d(0)$ is the initial mass of the disk, $\alpha$ is the dimensionless viscosity parameter and $T_{100}$ is the disk temperature at 100 AU.  The parameter $t_d$ is related to the age of the disk through,
\be
t_d=1+\frac{t}{t_s}, 
\en
where the viscous time, $t_s$ is given by,
\be
\begin{split}
t_s\sim&8\times10^4\;\left(\frac{R_1}{10\;\rm{AU}}\right)\left(\frac{\alpha}{10^{-2}}\right)^{-1}\left(\frac{M_{\ast}}{0.5\;\msun}\right)^{1/2}\\
&\times\left(\frac{T_{100}}{10\;\rm{K}}\right)^{-1}\;\rm{yr}.
\end{split}
\en
The quantity $R_1$ is the radius at which 60\% of the mass resides initially.  The fiducial viscous evolution model of \citet{hartmann98} is shown in Figure \ref{mdotage} as the thick dashed blue line, with parameters $M_d(0)=0.1\;\msun$, $\alpha=10^{-2}$, $M_\ast=0.5\;\msun$, $R_{\ast}=2\;\rsun$, $R_1=10$ AU and $T_{100}$=10 K.  This model assumes a similarity solution for disk evolution where the viscosity varies with radius in the disk but is constant in time.  Angular momentum is transported by viscous stresses and the disk expands to conserve angular momentum.  For a detailed description of the model see \citet{hartmann98}.

Given the variables in Equation \ref{h98}, there are several disk or stellar properties which may extend the accretion lifetime.  Increasing the initial disk mass achieves the desired effect as $\mdot \propto M_d(0)$; however, at $M_d(0)=0.1\; \msun$, the disk mass is nearing a critical point where it is too large relative to the mass of the star (in the range of 0.4 and 0.8 $\msun$ for the Orion our sample), causing the disk to be gravitationally unstable \citep{pringle81,larson84,gammie01}.  However, detailed analysis of the spectral energy distribution (SED) of young disks
indicates that in some cases $\alpha$ can be lower than
in the fiducial model.  \citet{mcclure13b} modeled the IR spectra of four CTTS in Taurus using the \citet{dalessio06} disk models and found values of $\alpha$ between $8\times10^{-4}$ and 0.05, therefore low $\alpha$ values may contribute to the long disk lifetime. 

In addition, circumstellar disk properties depend on the initial conditions of the cloud core which collapses to form the star and disk.  \citet{bae13} reproduced observed disk frequencies for a given age by assuming objects form from clouds with a distribution of angular momenta.  The angular velocity of the cloud core  ($\Omega$) determines where the mass is deposited in the disk, so it is related to $R_1$.  We assume $R_1$ may be approximated by the centrifugal radius ($R_c$), $R_1\sim R_c\propto\Omega^2$ \citep{cassen81}.  This means that we expect to find a distribution 
of $R_1$ consistent with the distribution of
cloud angular momenta.  Angular velocities derived from the velocity gradients observed in NH$_3$ cores range between $5\times10^{-15}$ and $5\times10^{-13}$ rad s$^{-1}$ \citep{goodman93}.   A wider distribution was predicted by \citet{dib10}, who used simulations of magnetized, self-gravitating molecular clouds.  They found that the distribution of specific angular momentum (core angular momentum $J$ divided by core mass ($M$)  peaks near log($J/M$)=20 cm$^2$ s$^{-1}$, corresponding to $\Omega=1\times10^{-15}$ rad s$^{-1}$, assuming uniform rotation at 0.1 pc, but can extend down to $10^{-16}$ rad s$^{-1}$.  

Figure \ref{mdotage} explores viscous evolution of $\mdot$ over the parameter space in $\alpha$ and $\Omega$, for $\alpha$=$10^{-3}$ and $10^{-2}$ and $\Omega=10^{-16} - 10^{-14}$ rad s$^{-1}$.  Each type of line (solid, dashed or dotted) shows constant $\Omega$, while the color and line weight (thin black or thick blue) represents constant $\alpha$.  The behavior of $\mdot$ with age may be understood by considering two cases: (1) when the viscous timescale is short compared to the age of the disk and (2) when it is the same or greater than the disk age.  In the first case, $t_d$ goes to $t/t_s$ and $\mdot \propto (R_1/\alpha)^{1/2} \propto \Omega/ \alpha^{1/2}$.  This condition is applicable to viscous evolution with low $\Omega$, shown by the solid and dashed lines in Figure \ref{mdotage}.  In the second case, for long viscous time scales, $\mdot \propto R_1^{-1} \alpha$ or $\mdot \propto \Omega^{-2} \alpha$.  At early times in Figure \ref{mdotage}, where $t_s>t$ (dotted lines, where $\Omega$ is large), the evolution of $\mdot$ follows this relation.  Eventually, $t$ increases until the age of the disk is larger than even long viscous timescales and the relation for $t<t_s$, takes over and describes the remaining evolution.

Compared to measured accretion rates, we find that the models still cannot explain high $\mdot$ objects in the Orion regions.  In fact, the simple models cannot explain even the highest accretors in Taurus, at 1 Myr either.  In order to fit the measured accretion rates, it is necessary to increase $M_d(0)$ to 0.5 $\msun$, 5$\times$ higher than the fiducial model.  It is likely that these high disk masses are gravitationally unstable.  Up to the epoch of observation, most objects have accreted $<$0.1 $\msun$, which is estimated by assuming each object has been accreting at the measured rate from $t=0$ to its present age.  The orange dash-dotted line in Figure \ref{mdotage} shows the upper limit on $\mdot$ where $\le$ 0.1 $\msun$ has been accreted over the source lifetime.  

A few explanations of the discrepancy between the measured and predicted $\mdot$ include time variability of accretion and the inapplicability of the similarity solution at large $t_s$.  T Tauri stars are known to be variable, and significant changes in brightness, up to 3 mag, have been observed on timescales of hours to weeks, with the shortest brightening events attributed to accretion \citep{herbst94}.  Therefore, high states of accretion may be responsible for some of the objects that cannot be described by viscous evolution, though it is unlikely that over half the objects in this analysis were observed during these short bursts of accretion.  Additionally, Equations 2--4 assume that $t_s$ is short enough that any initial conditions in the disk are quickly erased and further evolution may be described by the similarity solution.  If, however, the viscous timescale is long, initial conditions are important and the above equations are not valid.  Long viscous timescales may be necessary to explain the remaining high $\mdot$ objects at 10 Myr, and therefore we cannot assume the similarity solution.  Initial conditions like dead zones, which are not accounted for in the analysis so far, may provide an additional reservoir of material available in the disk.

 Dead zones are regions in the disk mid-plane which are not accreting \citep{gammie96}.  Dead zones have low (or 0) values of $\alpha$ and therefore have long viscous timescales.  For $\alpha=10^{-5}$,  $t_s > 10$ Myr, allowing the dead zone to retain its mass.  The maximum mass of the dead zone which may be stable against gravitational collapse is found by assuming the Toomre stability criterion is met,
 \be
 \label{toomre}
 Q=\frac{c_s\Omega_k}{\pi G \Sigma}<1.4,
 \en
 where $c_s$ is the sound speed, $\Omega_k$ is the orbital frequency and $\Sigma$ is the surface density in the disk.  The mass inside a dead zone extending to 10 AU is, 
 \be
 M_d=\int_0^{10\; \rm{AU}} \Sigma\; 2\pi R\; dR.
 \en
 Using the minimum $\Sigma$ for the dead zone to be stable (Equation \ref{toomre}) and assuming that the temperature scales as $T(R)=300\; \rm{K}\; (1\; \rm{AU}/$R$)^{1/2}$, from irradiated disk models at radii $>>$ than the stellar radius \citep{dalessio98,dalessio99,dalessio01}, the maximum mass available in the dead zone is 0.2 $\msun$.  This provides a significant amount of material available for accretion at later stages.  
 
 In addition to viscous evolution, there are other processes which may deplete the disk and shorten the accretion lifetime, in particular photoevaporation and planet formation, which we address in the next section.

\subsection{Photoevaporation and planet formation in the case of CVSO 206}

\begin{figure}[htp]
\plotone{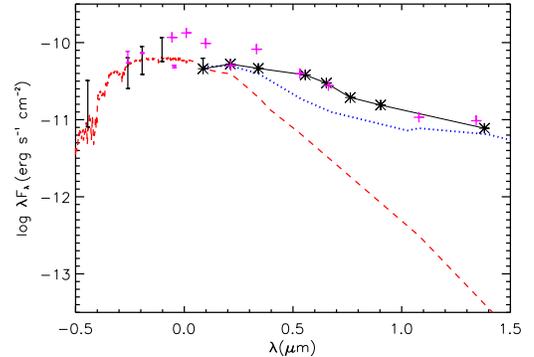}
\caption{SED of CVSO 206.  The range of MDM $U, V, R, I$ photometry (black bars), taken over a 4 night span is plotted along with 2MASS $JHK$, IRAC and MIPS photometry from \citet{hernandez07a}, shown as black asterisks.  A K7 photosphere is shown (red dashed) and the median SED of Taurus is plotted as well \citep[blue dotted]{dalessio99}, both normalized at $J$.  Magenta points represent non-simultaneous optical data from \citet{briceno05} as well as $Z,Y,J,H,K$ data from VISTA (Visible and Infrared Survey Telescope for Astronomy) and IR photometry from WISE (Wide-field Infrared Survey Explorer).  The IR emission from CVSO 206 is in clear excess over the median of Taurus, indicating that it retains a full disk.}
\label{sed}
\end{figure}

Significant evolution of the circumstellar disk is likely to happen in the first 10 Myr, so objects in Orion OB1a are expected to show evidence of this evolution.  We consider a member of this subassociation, CVSO 206, as an example of a 7--10 Myr object with significant ongoing accretion.  Figure \ref{sed} shows the SED of CVSO 206.  Compared to the median SED of all objects with disks in the Taurus star forming region, CVSO 206 has a significant excess, showing that it retains a full dust disk.  Gas is not probed by the IR excess; however, given the accretion rate it is likely that the CVSO 206 has a full gas disk as well.    With an accretion rate of $1\times10^{-8}\;\msunyr$, the star would have accreted a minimum of 0.06--0.09 $\msun$ between 1 Myr and its current age.  Therefore, the minimum mass of the disk around CVSO 206 when it was 1 Myr was greater, $M_d(1\;\rm{Myr})>0.06-0.09\;\msun$.  With $M_{\ast}=0.8\;\msun$, the disk to star mass ratio of this object at 1 Myr would have been $M_d(1\;\rm{Myr})/M_{\ast}>$ 0.0875--0.125, or $\sim$10\%.   Typical disk masses in Taurus are in the range of 0.2 - 0.6\% of the stellar mass \citep{andrews13}, indicating that CVSO 206 had an atypically massive disk at 1 Myr.

Many processes attempt to deplete the disk; including photoevaporation by high energy radiation from the central star or nearby hot stars and planet formation.  Photoevaporation by the central star does not become a factor until the mass accretion rate onto the star drops below the mass loss rate.  According to photoevaporation models, once the mass accretion and mass loss rates are comparable, a gap will open in the circumstellar disk, after which the inner disk accretes quickly onto the central star.  After the inner disk is gone, the outer disk is directly exposed to radiation from the star and is itself depleted quickly, all in $\sim10^5$ yr.  For some models this event may occur earlier than others; for instance X-ray and FUV photoevaporation models predict mass loss rates of $10^{-8}\;\msunyr$ \citep{owen10,gorti09} while in EUV photoevaporation models the mass loss occurs at lower rates of $10^{-10}\;\msunyr$ \citep{clarke01,alexander06}.    The fact that there are a number of sources with accretion rates $<10^{-8}\;\msunyr$ in the 5 and 10 Myr Orion OB1b and OB1a regions lends more support to the low mass loss rates.

External photoevaporation by FUV radiation from nearby OB stars is another disk dispersal mechanism.  Massive stars drive mass loss from the outer radii of nearby circumstellar disks, dispersing the disk mass in short timescales for the closest disks.  \citet{anderson13} showed that even if external FUV radiation is low the disk is quickly truncated to 100 AU or smaller.   However, combining low values of $G_0$ (ratio of UV radiation in a cluster to the typical value of the interstellar medium) with low $\alpha$ results in a circumstellar disk which may be sustained against external photoevaporation.  The group 25 Ori, where CVSO 206 is located, has a population of $N\sim$ 500 stars, assuming a standard initial mass function (IMF) covering 7 pc \citep{briceno05}.  According to results of \citet{fatuzzo08}, the mean FUV luminosity of a cluster with 500 stars is $\sim5\times10^{37}$ erg s$^{-1}$.  Assuming the Be star 25 Orionis is at the center of the cluster, the projected distance of  CVSO 206 from the center is 0.4 pc.  The external FUV flux at this distance, compared to the typical interstellar value of 1.6$\times10^{-3}$ erg s$^{-1}$ cm$^{-2}$, gives $G_0=1630$.  For the disk to have survived this degree of external radiation, it must have a very low $\alpha$ value.

Dust grain coagulation, while not depleting the disk, decreases the number of small grains responsible for producing the IR excess \citep{dullemond05}.   The growth of grains to planetary sizes is expected to occur between 1 and 10 Myr \citep{pollack96} and planets appear to form frequently, with over 1000 planetary candidates identified in a recent Kepler release \citep{batalha13}.   Massive planets may form gaps or holes in the dust distribution by sweeping up material as they orbit  \citep{lin86,bryden99,calvet02,espaillat10}.   \citet{zhu11} showed that it may require multiple planets to open a gap the size of which can be identified in infrared spectra.   CVSO 206 shows no evidence for a gap or hole in the disk and must have remaining small grains to produce the observed IR excess.  Still, it may contain planets which are not massive enough to open a detectable gap.

In summary, to explain the remaining full disk around CVSO 206, the disk may have low viscosity, slowing the spread to large radii and diminishing the effect of external radiation driving mass loss from the disk.  Given that lower $\alpha$ values increase the accretion lifetime and therefore may be characteristic of this disk, it is feasible that the disk of CVSO 206 could survive against external photoevaporation as long as 10 Myr \citep{anderson13}.  With $\mdot=1\times10^{-8}\;\msunyr$, photoevaporation by the central object may be dominated by any of the X-ray, FUV or EUV fields as the accretion rate still exceeds the mass loss rates predicted by all models.   As a result, we would not expect internal photoevaporation to have significantly altered the disk.  Finally, the IR SED shows that planets the size or quantity capable of opening a large gap in the disk must not be present, though small planets in the disk may not be ruled out.  \citet{birnstiel09} showed that collisions causing fragmentation of large grains in the disk are capable of replenishing the supply of small grains which are the main opacity source at near-IR wavelengths.  This may be an indication that collisions which have contributed to the remainder of small dust grains have halted efficient planet formation.

\section{Summary}
We used optical spectra along with optical and ultraviolet photometry to estimate mass accretion rates for objects in the distributed star forming region around Orion, including the $\sigma$ Ori cluster, Orion OB1b and the 25 Ori group within Orion OB1a.  The following conclusions were drawn from our analysis.

\begin{itemize}
\item
Previous observations of older CTTS, in the 5--10 Myr range, indicate that accretion slows with the age of the object; however, samples of evolved CTTS are small.  Here, we showed several examples of objects as old as 10 Myr with ongoing significant accretion rates.   These objects have values of $\mdot$ similar to objects in the Taurus star forming region. 

\item
We compared the accretion rates of our objects to the predicted decline in $\mdot$ due to viscous evolution.  We found that simple viscous evolution models, which take into account a range of viscosities and disk outer radii, cannot explain high accretors at either 1 Myr or 10 Myr, unless the initial disk mass is very high.  The disk masses needed would cause the disk to be gravitationally unstable.  Time variability, or accretion high states, may explain some of the discrepancy.  Additionally, the similarity solutions we assume to describe the viscous evolution are not applicable when the viscous timescale is comparable to or longer than the age of the system, a condition that is likely needed for prolonged accretion.

\item
CVSO 206 is a 7--10 Myr strong accretor, with $\mdot=7\times10^{-9}\;\msunyr$, indicating that it had a disk mass at 1 Myr $>$0.06 $\msun$, or $M_{d}(1\;\rm{Myr})$/$M_{\ast}\sim$0.1, significantly higher than the typical disk mass in Taurus.  In addition, CVSO 206 has IR fluxes similar to the median of Taurus, and therefore significant grain growth and massive or multiple planet formation has not occurred.  Photoevaporation by either the central star or external stars has not played a key role in the disk evolution, likely because the high mass accretion rate is capable of replenishing the inner disk material dispersed by radiation from the star. 

\end{itemize}

\section{Acknowledgments}


\begin{thebibliography}{98}
\expandafter\ifx\csname natexlab\endcsname\relax\def\natexlab#1{#1}\fi

\bibitem[{{Adams} {et~al.}(2004){Adams}, {Hollenbach}, {Laughlin}, \&
  {Gorti}}]{adams04}
{Adams}, F.~C., {Hollenbach}, D., {Laughlin}, G., \& {Gorti}, U. 2004, \apj,
  611, 360

\bibitem[{{Alexander} {et~al.}(2006){Alexander}, {Clarke}, \&
  {Pringle}}]{alexander06}
{Alexander}, R.~D., {Clarke}, C.~J., \& {Pringle}, J.~E. 2006, \mnras, 369, 229

\bibitem[{{Anderson} {et~al.}(2013){Anderson}, {Adams}, \&
  {Calvet}}]{anderson13}
{Anderson}, K.~R., {Adams}, F.~C., \& {Calvet}, N. 2013, ArXiv e-prints

\bibitem[{{Andrews} {et~al.}(2013){Andrews}, {Rosenfeld}, {Kraus}, \&
  {Wilner}}]{andrews13}
{Andrews}, S.~M., {Rosenfeld}, K.~A., {Kraus}, A.~L., \& {Wilner}, D.~J. 2013,
  \apj, 771, 129

\bibitem[{{Bae} {et~al.}(2013){Bae}, {Hartmann}, {Zhu}, \& {Gammie}}]{bae13}
{Bae}, J., {Hartmann}, L., {Zhu}, Z., \& {Gammie}, C. 2013, \apj, 774, 57

\bibitem[{{Barrado y Navascu{\'e}s} \& {Mart{\'{\i}}n}(2003)}]{barrado03}
{Barrado y Navascu{\'e}s}, D., \& {Mart{\'{\i}}n}, E.~L. 2003, \aj, 126, 2997

\bibitem[{{Batalha} {et~al.}(2013){Batalha}, {Rowe}, {Bryson}, {Barclay},
  {Burke}, {Caldwell}, {Christiansen}, {Mullally}, {Thompson}, {Brown},
  {Dupree}, {Fabrycky}, {Ford}, {Fortney}, {Gilliland}, {Isaacson}, {Latham},
  {Marcy}, {Quinn}, {Ragozzine}, {Shporer}, {Borucki}, {Ciardi}, {Gautier},
  {Haas}, {Jenkins}, {Koch}, {Lissauer}, {Rapin}, {Basri}, {Boss}, {Buchhave},
  {Carter}, {Charbonneau}, {Christensen-Dalsgaard}, {Clarke}, {Cochran},
  {Demory}, {Desert}, {Devore}, {Doyle}, {Esquerdo}, {Everett}, {Fressin},
  {Geary}, {Girouard}, {Gould}, {Hall}, {Holman}, {Howard}, {Howell},
  {Ibrahim}, {Kinemuchi}, {Kjeldsen}, {Klaus}, {Li}, {Lucas}, {Meibom},
  {Morris}, {Pr{\v s}a}, {Quintana}, {Sanderfer}, {Sasselov}, {Seader},
  {Smith}, {Steffen}, {Still}, {Stumpe}, {Tarter}, {Tenenbaum}, {Torres},
  {Twicken}, {Uddin}, {Van Cleve}, {Walkowicz}, \& {Welsh}}]{batalha13}
{Batalha}, N.~M., {Rowe}, J.~F., {Bryson}, S.~T., {et~al.} 2013, \apjs, 204, 24

\bibitem[{{Birnstiel} {et~al.}(2009){Birnstiel}, {Dullemond}, \&
  {Brauer}}]{birnstiel09}
{Birnstiel}, T., {Dullemond}, C.~P., \& {Brauer}, F. 2009, \aap, 503, L5

\bibitem[{{Bouvier} {et~al.}(2007){Bouvier}, {Alencar}, {Harries},
  {Johns-Krull}, \& {Romanova}}]{bouvier07b}
{Bouvier}, J., {Alencar}, S.~H.~P., {Harries}, T.~J., {Johns-Krull}, C.~M., \&
  {Romanova}, M.~M. 2007, Protostars and Planets V, 479

\bibitem[{{Brice{\~n}o} {et~al.}(2005){Brice{\~n}o}, {Calvet}, {Hern{\'a}ndez},
  {Vivas}, {Hartmann}, {Downes}, \& {Berlind}}]{briceno05}
{Brice{\~n}o}, C., {Calvet}, N., {Hern{\'a}ndez}, J., {et~al.} 2005, \aj, 129,
  907

\bibitem[{{Brice{\~n}o} {et~al.}(2007){Brice{\~n}o}, {Hartmann},
  {Hern{\'a}ndez}, {Calvet}, {Vivas}, {Furesz}, \& {Szentgyorgyi}}]{briceno07}
{Brice{\~n}o}, C., {Hartmann}, L., {Hern{\'a}ndez}, J., {et~al.} 2007, \apj,
  661, 1119

\bibitem[{{Bryden} {et~al.}(1999){Bryden}, {Chen}, {Lin}, {Nelson}, \&
  {Papaloizou}}]{bryden99}
{Bryden}, G., {Chen}, X., {Lin}, D.~N.~C., {Nelson}, R.~P., \& {Papaloizou},
  J.~C.~B. 1999, \apj, 514, 344

\bibitem[{{Caballero}(2008)}]{caballero08}
{Caballero}, J.~A. 2008, \aap, 478, 667

\bibitem[{{Calvet} {et~al.}(2005){Calvet}, {Brice{\~n}o}, {Hern{\'a}ndez},
  {Hoyer}, {Hartmann}, {Sicilia-Aguilar}, {Megeath}, \& {D'Alessio}}]{calvet05}
{Calvet}, N., {Brice{\~n}o}, C., {Hern{\'a}ndez}, J., {et~al.} 2005, \aj, 129,
  935

\bibitem[{{Calvet} {et~al.}(2002){Calvet}, {D'Alessio}, {Hartmann}, {Wilner},
  {Walsh}, \& {Sitko}}]{calvet02}
{Calvet}, N., {D'Alessio}, P., {Hartmann}, L., {et~al.} 2002, \apj, 568, 1008

\bibitem[{{Calvet} \& {Gullbring}(1998)}]{calvet98}
{Calvet}, N., \& {Gullbring}, E. 1998, \apj, 509, 802

\bibitem[{{Cassen} \& {Moosman}(1981)}]{cassen81}
{Cassen}, P., \& {Moosman}, A. 1981, \icarus, 48, 353

\bibitem[{{Clarke} {et~al.}(2001){Clarke}, {Gendrin}, \&
  {Sotomayor}}]{clarke01}
{Clarke}, C.~J., {Gendrin}, A., \& {Sotomayor}, M. 2001, \mnras, 328, 485

\bibitem[{{D'Alessio} {et~al.}(2001){D'Alessio}, {Calvet}, \&
  {Hartmann}}]{dalessio01}
{D'Alessio}, P., {Calvet}, N., \& {Hartmann}, L. 2001, \apj, 553, 321

\bibitem[{{D'Alessio} {et~al.}(2006){D'Alessio}, {Calvet}, {Hartmann},
  {Franco-Hern{\'a}ndez}, \& {Serv{\'{\i}}n}}]{dalessio06}
{D'Alessio}, P., {Calvet}, N., {Hartmann}, L., {Franco-Hern{\'a}ndez}, R., \&
  {Serv{\'{\i}}n}, H. 2006, \apj, 638, 314

\bibitem[{{D'Alessio} {et~al.}(1999){D'Alessio}, {Calvet}, {Hartmann},
  {Lizano}, \& {Cant{\'o}}}]{dalessio99}
{D'Alessio}, P., {Calvet}, N., {Hartmann}, L., {Lizano}, S., \& {Cant{\'o}}, J.
  1999, \apj, 527, 893

\bibitem[{{D'Alessio} {et~al.}(1998){D'Alessio}, {Canto}, {Calvet}, \&
  {Lizano}}]{dalessio98}
{D'Alessio}, P., {Canto}, J., {Calvet}, N., \& {Lizano}, S. 1998, \apj, 500,
  411

\bibitem[{{Dib} {et~al.}(2010){Dib}, {Hennebelle}, {Pineda}, {Csengeri},
  {Bontemps}, {Audit}, \& {Goodman}}]{dib10}
{Dib}, S., {Hennebelle}, P., {Pineda}, J.~E., {et~al.} 2010, \apj, 723, 425

\bibitem[{{Dodin} \& {Lamzin}(2012)}]{dodin12}
{Dodin}, A.~V., \& {Lamzin}, S.~A. 2012, ArXiv e-prints

\bibitem[{{Donati} {et~al.}(2008){Donati}, {Jardine}, {Gregory}, {Petit},
  {Paletou}, {Bouvier}, {Dougados}, {M{\'e}nard}, {Collier Cameron}, {Harries},
  {Hussain}, {Unruh}, {Morin}, {Marsden}, {Manset}, {Auri{\`e}re}, {Catala}, \&
  {Alecian}}]{donati08}
{Donati}, J.-F., {Jardine}, M.~M., {Gregory}, S.~G., {et~al.} 2008, \mnras,
  386, 1234

\bibitem[{{Dullemond} \& {Dominik}(2005)}]{dullemond05}
{Dullemond}, C.~P., \& {Dominik}, C. 2005, \aap, 434, 971

\bibitem[{{Edwards} {et~al.}(2006){Edwards}, {Fischer}, {Hillenbrand}, \&
  {Kwan}}]{edwards06}
{Edwards}, S., {Fischer}, W., {Hillenbrand}, L., \& {Kwan}, J. 2006, \apj, 646,
  319

\bibitem[{{Edwards} {et~al.}(1994){Edwards}, {Hartigan}, {Ghandour}, \&
  {Andrulis}}]{edwards94}
{Edwards}, S., {Hartigan}, P., {Ghandour}, L., \& {Andrulis}, C. 1994, \aj,
  108, 1056

\bibitem[{{Espaillat} {et~al.}(2008){Espaillat}, {Muzerolle}, {Hern{\'a}ndez},
  {Brice{\~n}o}, {Calvet}, {D'Alessio}, {McClure}, {Watson}, {Hartmann}, \&
  {Sargent}}]{espaillat08b}
{Espaillat}, C., {Muzerolle}, J., {Hern{\'a}ndez}, J., {et~al.} 2008, \apjl,
  689, L145

\bibitem[{{Espaillat} {et~al.}(2010){Espaillat}, {D'Alessio}, {Hern{\'a}ndez},
  {Nagel}, {Luhman}, {Watson}, {Calvet}, {Muzerolle}, \&
  {McClure}}]{espaillat10}
{Espaillat}, C., {D'Alessio}, P., {Hern{\'a}ndez}, J., {et~al.} 2010, \apj,
  717, 441

\bibitem[{{Fatuzzo} \& {Adams}(2008)}]{fatuzzo08}
{Fatuzzo}, M., \& {Adams}, F.~C. 2008, \apj, 675, 1361

\bibitem[{{Fedele} {et~al.}(2010){Fedele}, {van den Ancker}, {Henning},
  {Jayawardhana}, \& {Oliveira}}]{fedele10}
{Fedele}, D., {van den Ancker}, M.~E., {Henning}, T., {Jayawardhana}, R., \&
  {Oliveira}, J.~M. 2010, \aap, 510, A72

\bibitem[{{Fischer} {et~al.}(2011){Fischer}, {Edwards}, {Hillenbrand}, \&
  {Kwan}}]{fischer11}
{Fischer}, W., {Edwards}, S., {Hillenbrand}, L., \& {Kwan}, J. 2011, \apj, 730,
  73

\bibitem[{{Font} {et~al.}(2004){Font}, {McCarthy}, {Johnstone}, \&
  {Ballantyne}}]{font04}
{Font}, A.~S., {McCarthy}, I.~G., {Johnstone}, D., \& {Ballantyne}, D.~R. 2004,
  \apj, 607, 890

\bibitem[{{France} {et~al.}(2012){France}, {Schindhelm}, {Herczeg}, {Brown},
  {Abgrall}, {Alexander}, {Bergin}, {Brown}, {Linsky}, {Roueff}, \&
  {Yang}}]{france12b}
{France}, K., {Schindhelm}, E., {Herczeg}, G.~J., {et~al.} 2012, \apj, 756, 171

\bibitem[{{Gahm} {et~al.}(2008){Gahm}, {Walter}, {Stempels}, {Petrov}, \&
  {Herczeg}}]{gahm08}
{Gahm}, G.~F., {Walter}, F.~M., {Stempels}, H.~C., {Petrov}, P.~P., \&
  {Herczeg}, G.~J. 2008, \aap, 482, L35

\bibitem[{{Gammie}(1996)}]{gammie96}
{Gammie}, C.~F. 1996, \apj, 457, 355

\bibitem[{{Gammie}(2001)}]{gammie01}
---. 2001, \apj, 553, 174

\bibitem[{{Goodman} {et~al.}(1993){Goodman}, {Benson}, {Fuller}, \&
  {Myers}}]{goodman93}
{Goodman}, A.~A., {Benson}, P.~J., {Fuller}, G.~A., \& {Myers}, P.~C. 1993,
  \apj, 406, 528

\bibitem[{{Gorti} {et~al.}(2009){Gorti}, {Dullemond}, \&
  {Hollenbach}}]{gorti09}
{Gorti}, U., {Dullemond}, C.~P., \& {Hollenbach}, D. 2009, \apj, 705, 1237

\bibitem[{{Gregory} \& {Donati}(2011)}]{gregory11}
{Gregory}, S.~G., \& {Donati}, J.-F. 2011, Astronomische Nachrichten, 332, 1027

\bibitem[{{Gregory} {et~al.}(2012){Gregory}, {Donati}, {Morin}, {Hussain},
  {Mayne}, {Hillenbrand}, \& {Jardine}}]{gregory12}
{Gregory}, S.~G., {Donati}, J.-F., {Morin}, J., {et~al.} 2012, \apj, 755, 97

\bibitem[{{Guenther} {et~al.}(2013){Guenther}, {Wolter}, {Robrade}, \&
  {Wolk}}]{guenther13}
{Guenther}, H.~M., {Wolter}, U., {Robrade}, J., \& {Wolk}, S.~J. 2013, ArXiv
  e-prints

\bibitem[{{Gullbring} {et~al.}(1998){Gullbring}, {Hartmann}, {Briceno}, \&
  {Calvet}}]{gullbring98}
{Gullbring}, E., {Hartmann}, L., {Briceno}, C., \& {Calvet}, N. 1998, \apj,
  492, 323

\bibitem[{{Hartigan} {et~al.}(1989){Hartigan}, {Hartmann}, {Kenyon}, {Hewett},
  \& {Stauffer}}]{hartigan89}
{Hartigan}, P., {Hartmann}, L., {Kenyon}, S., {Hewett}, R., \& {Stauffer}, J.
  1989, \apjs, 70, 899

\bibitem[{{Hartmann}(2009)}]{hartmann09}
{Hartmann}, L. 2009, {Accretion Processes in Star Formation: Second Edition}
  (Cambridge University Press)

\bibitem[{{Hartmann} {et~al.}(1998){Hartmann}, {Calvet}, {Gullbring}, \&
  {D'Alessio}}]{hartmann98}
{Hartmann}, L., {Calvet}, N., {Gullbring}, E., \& {D'Alessio}, P. 1998, \apj,
  495, 385

\bibitem[{{Hartmann} {et~al.}(1994){Hartmann}, {Hewett}, \&
  {Calvet}}]{hartmann94}
{Hartmann}, L., {Hewett}, R., \& {Calvet}, N. 1994, \apj, 426, 669

\bibitem[{{Hartmann} \& {Kenyon}(1985)}]{hartmann85}
{Hartmann}, L., \& {Kenyon}, S.~J. 1985, \apj, 299, 462

\bibitem[{{Herbig} \& {Bell}(1988)}]{herbig88}
{Herbig}, G.~H., \& {Bell}, K.~R. 1988, {Third Catalog of Emission-Line Stars
  of the Orion Population : 3 : 1988}

\bibitem[{{Herbst} {et~al.}(1994){Herbst}, {Herbst}, {Grossman}, \&
  {Weinstein}}]{herbst94}
{Herbst}, W., {Herbst}, D.~K., {Grossman}, E.~J., \& {Weinstein}, D. 1994, \aj,
  108, 1906

\bibitem[{{Herczeg} \& {Hillenbrand}(2008)}]{herczeg08}
{Herczeg}, G.~J., \& {Hillenbrand}, L.~A. 2008, \apj, 681, 594

\bibitem[{{Hern{\'a}ndez} {et~al.}(2004){Hern{\'a}ndez}, {Calvet},
  {Brice{\~n}o}, {Hartmann}, \& {Berlind}}]{hernandez04}
{Hern{\'a}ndez}, J., {Calvet}, N., {Brice{\~n}o}, C., {Hartmann}, L., \&
  {Berlind}, P. 2004, \aj, 127, 1682

\bibitem[{{Hern{\'a}ndez} {et~al.}(2008){Hern{\'a}ndez}, {Hartmann}, {Calvet},
  {Jeffries}, {Gutermuth}, {Muzerolle}, \& {Stauffer}}]{hernandez08}
{Hern{\'a}ndez}, J., {Hartmann}, L., {Calvet}, N., {et~al.} 2008, \apj, 686,
  1195

\bibitem[{{Hern{\'a}ndez} {et~al.}(2007{\natexlab{a}}){Hern{\'a}ndez},
  {Hartmann}, {Megeath}, {Gutermuth}, {Muzerolle}, {Calvet}, {Vivas},
  {Brice{\~n}o}, {Allen}, {Stauffer}, {Young}, \& {Fazio}}]{hernandez07b}
{Hern{\'a}ndez}, J., {Hartmann}, L., {Megeath}, T., {et~al.}
  2007{\natexlab{a}}, \apj, 662, 1067

\bibitem[{{Hern{\'a}ndez} {et~al.}(2007{\natexlab{b}}){Hern{\'a}ndez},
  {Calvet}, {Brice{\~n}o}, {Hartmann}, {Vivas}, {Muzerolle}, {Downes}, {Allen},
  \& {Gutermuth}}]{hernandez07a}
{Hern{\'a}ndez}, J., {Calvet}, N., {Brice{\~n}o}, C., {et~al.}
  2007{\natexlab{b}}, \apj, 671, 1784

\bibitem[{{Hsu} {et~al.}(2012){Hsu}, {Hartmann}, {Allen}, {Hern{\'a}ndez},
  {Megeath}, {Mosby}, {Tobin}, \& {Espaillat}}]{hsu12}
{Hsu}, W.-H., {Hartmann}, L., {Allen}, L., {et~al.} 2012, \apj, 752, 59

\bibitem[{{Ingleby} {et~al.}(2012){Ingleby}, {Calvet}, {Herczeg}, \&
  {Brice{\~n}o}}]{ingleby12}
{Ingleby}, L., {Calvet}, N., {Herczeg}, G., \& {Brice{\~n}o}, C. 2012, \apjl,
  752, L20

\bibitem[{{Ingleby} {et~al.}(2011){Ingleby}, {Calvet}, {Bergin}, {Herczeg},
  {Brown}, {Alexander}, {Edwards}, {Espaillat}, {France}, {Gregory},
  {Hillenbrand}, {Roueff}, {Valenti}, {Walter}, {Johns-Krull}, {Brown},
  {Linsky}, {McClure}, {Ardila}, {Abgrall}, {Bethell}, {Hussain}, \&
  {Yang}}]{ingleby11b}
{Ingleby}, L., {Calvet}, N., {Bergin}, E., {et~al.} 2011, \apj, 743, 105

\bibitem[{{Ingleby} {et~al.}(2013){Ingleby}, {Calvet}, {Herczeg}, {Blaty},
  {Walter}, {Ardila}, {Alexander}, {Edwards}, {Espaillat}, {Gregory},
  {Hillenbrand}, \& {Brown}}]{ingleby13}
{Ingleby}, L., {Calvet}, N., {Herczeg}, G., {et~al.} 2013, \apj, 767, 112

\bibitem[{{Kenyon} \& {Hartmann}(1995)}]{kenyon95}
{Kenyon}, S.~J., \& {Hartmann}, L. 1995, \apjs, 101, 117

\bibitem[{{Kretke} \& {Lin}(2012)}]{kretke12}
{Kretke}, K.~A., \& {Lin}, D.~N.~C. 2012, \apj, 755, 74

\bibitem[{{Larson}(1984)}]{larson84}
{Larson}, R.~B. 1984, \mnras, 206, 197

\bibitem[{{Lin} \& {Papaloizou}(1986)}]{lin86}
{Lin}, D.~N.~C., \& {Papaloizou}, J. 1986, \apj, 309, 846

\bibitem[{{Long} {et~al.}(2011){Long}, {Romanova}, {Kulkarni}, \&
  {Donati}}]{long11}
{Long}, M., {Romanova}, M.~M., {Kulkarni}, A.~K., \& {Donati}, J.-F. 2011,
  \mnras, 413, 1061

\bibitem[{{Lubow} \& {D'Angelo}(2006)}]{lubow06}
{Lubow}, S.~H., \& {D'Angelo}, G. 2006, \apj, 641, 526

\bibitem[{{Luhman}(2004)}]{luhman04}
{Luhman}, K.~L. 2004, \apj, 602, 816

\bibitem[{{Manara} {et~al.}(2013){Manara}, {Beccari}, {Da Rio}, {De Marchi},
  {Natta}, {Ricci}, {Robberto}, \& {Testi}}]{manara13}
{Manara}, C.~F., {Beccari}, G., {Da Rio}, N., {et~al.} 2013, \aap, 558, A114

\bibitem[{{McClure} {et~al.}(2013{\natexlab{a}}){McClure}, {Calvet},
  {Espaillat}, {Hartmann}, {Hern{\'a}ndez}, {Ingleby}, {Luhman}, {D'Alessio},
  \& {Sargent}}]{mcclure13a}
{McClure}, M.~K., {Calvet}, N., {Espaillat}, C., {et~al.} 2013{\natexlab{a}},
  \apj, 769, 73

\bibitem[{{McClure} {et~al.}(2013{\natexlab{b}}){McClure}, {D'Alessio},
  {Calvet}, {Espaillat}, {Hartmann}, {Sargent}, {Watson}, {Ingleby}, \&
  {Hern{\'a}ndez}}]{mcclure13b}
{McClure}, M.~K., {D'Alessio}, P., {Calvet}, N., {et~al.} 2013{\natexlab{b}},
  \apj, 775, 114

\bibitem[{{Merle} {et~al.}(2011){Merle}, {Th{\'e}venin}, {Pichon}, \&
  {Bigot}}]{merle11}
{Merle}, T., {Th{\'e}venin}, F., {Pichon}, B., \& {Bigot}, L. 2011, \mnras,
  418, 863

\bibitem[{{Muzerolle} {et~al.}(2000){Muzerolle}, {Calvet}, {Brice{\~n}o},
  {Hartmann}, \& {Hillenbrand}}]{muzerolle00}
{Muzerolle}, J., {Calvet}, N., {Brice{\~n}o}, C., {Hartmann}, L., \&
  {Hillenbrand}, L. 2000, \apjl, 535, L47

\bibitem[{{Muzerolle} {et~al.}(1998){Muzerolle}, {Calvet}, \&
  {Hartmann}}]{muzerolle98}
{Muzerolle}, J., {Calvet}, N., \& {Hartmann}, L. 1998, \apj, 492, 743

\bibitem[{{Muzerolle} {et~al.}(2001){Muzerolle}, {Calvet}, \&
  {Hartmann}}]{muzerolle01}
---. 2001, \apj, 550, 944

\bibitem[{{Muzerolle} {et~al.}(2005){Muzerolle}, {Luhman}, {Brice{\~n}o},
  {Hartmann}, \& {Calvet}}]{muzerolle05}
{Muzerolle}, J., {Luhman}, K.~L., {Brice{\~n}o}, C., {Hartmann}, L., \&
  {Calvet}, N. 2005, \apj, 625, 906

\bibitem[{{Natta} {et~al.}(2004){Natta}, {Testi}, {Muzerolle}, {Randich},
  {Comer{\'o}n}, \& {Persi}}]{natta04}
{Natta}, A., {Testi}, L., {Muzerolle}, J., {et~al.} 2004, \aap, 424, 603

\bibitem[{{Oliveira} {et~al.}(2002){Oliveira}, {Jeffries}, {Kenyon},
  {Thompson}, \& {Naylor}}]{oliveira02}
{Oliveira}, J.~M., {Jeffries}, R.~D., {Kenyon}, M.~J., {Thompson}, S.~A., \&
  {Naylor}, T. 2002, \aap, 382, L22

\bibitem[{{Owen} {et~al.}(2012{\natexlab{a}}){Owen}, {Clarke}, \&
  {Ercolano}}]{owen11}
{Owen}, J.~E., {Clarke}, C.~J., \& {Ercolano}, B. 2012{\natexlab{a}}, \mnras,
  422, 1880

\bibitem[{{Owen} {et~al.}(2012{\natexlab{b}}){Owen}, {Clarke}, \&
  {Ercolano}}]{owen12}
---. 2012{\natexlab{b}}, \mnras, 422, 1880

\bibitem[{{Owen} {et~al.}(2010){Owen}, {Ercolano}, {Clarke}, \&
  {Alexander}}]{owen10}
{Owen}, J.~E., {Ercolano}, B., {Clarke}, C.~J., \& {Alexander}, R.~D. 2010,
  \mnras, 401, 1415

\bibitem[{{Pecaut} \& {Mamajek}(2013)}]{pecaut13}
{Pecaut}, M.~J., \& {Mamajek}, E.~E. 2013, \apjs, 208, 9

\bibitem[{{Pollack} {et~al.}(1996){Pollack}, {Hubickyj}, {Bodenheimer},
  {Lissauer}, {Podolak}, \& {Greenzweig}}]{pollack96}
{Pollack}, J.~B., {Hubickyj}, O., {Bodenheimer}, P., {et~al.} 1996, \icarus,
  124, 62

\bibitem[{{Pringle}(1981)}]{pringle81}
{Pringle}, J.~E. 1981, \araa, 19, 137

\bibitem[{{Rigliaco} {et~al.}(2011){Rigliaco}, {Natta}, {Randich}, {Testi}, \&
  {Biazzo}}]{rigliaco11}
{Rigliaco}, E., {Natta}, A., {Randich}, S., {Testi}, L., \& {Biazzo}, K. 2011,
  \aap, 525, A47

\bibitem[{{Rosotti} {et~al.}(2013){Rosotti}, {Ercolano}, {Owen}, \&
  {Armitage}}]{rosotti13}
{Rosotti}, G.~P., {Ercolano}, B., {Owen}, J.~E., \& {Armitage}, P.~J. 2013,
  \mnras, 430, 1392

\bibitem[{{Sacco} {et~al.}(2008){Sacco}, {Argiroffi}, {Orlando}, {Maggio},
  {Peres}, \& {Reale}}]{sacco08}
{Sacco}, G.~G., {Argiroffi}, C., {Orlando}, S., {et~al.} 2008, \aap, 491, L17

\bibitem[{{Schindhelm} {et~al.}(2012){Schindhelm}, {France}, {Burgh},
  {Herczeg}, {Green}, {Brown}, {Brown}, \& {Valenti}}]{schindhelm12}
{Schindhelm}, E., {France}, K., {Burgh}, E.~B., {et~al.} 2012, \apj, 746, 97

\bibitem[{{Sherry} {et~al.}(2004){Sherry}, {Walter}, \& {Wolk}}]{sherry04}
{Sherry}, W.~H., {Walter}, F.~M., \& {Wolk}, S.~J. 2004, \aj, 128, 2316

\bibitem[{{Siess} {et~al.}(2000){Siess}, {Dufour}, \& {Forestini}}]{siess00}
{Siess}, L., {Dufour}, E., \& {Forestini}, M. 2000, \aap, 358, 593

\bibitem[{{Skrutskie} {et~al.}(2006){Skrutskie}, {Cutri}, {Stiening},
  {Weinberg}, {Schneider}, {Carpenter}, {Beichman}, {Capps}, {Chester},
  {Elias}, {Huchra}, {Liebert}, {Lonsdale}, {Monet}, {Price}, {Seitzer},
  {Jarrett}, {Kirkpatrick}, {Gizis}, {Howard}, {Evans}, {Fowler}, {Fullmer},
  {Hurt}, {Light}, {Kopan}, {Marsh}, {McCallon}, {Tam}, {Van Dyk}, \&
  {Wheelock}}]{skrutskie06}
{Skrutskie}, M.~F., {Cutri}, R.~M., {Stiening}, R., {et~al.} 2006, \aj, 131,
  1163

\bibitem[{{Tody}(1993)}]{tody93}
{Tody}, D. 1993, in Astronomical Society of the Pacific Conference Series,
  Vol.~52, Astronomical Data Analysis Software and Systems II, ed. R.~J.
  {Hanisch}, R.~J.~V. {Brissenden}, \& J.~{Barnes}, 173

\bibitem[{{Uchida} \& {Shibata}(1984)}]{uchida84}
{Uchida}, Y., \& {Shibata}, K. 1984, PASJ, 36, 105

\bibitem[{{Vorobyov} \& {Basu}(2006)}]{vorobyov06}
{Vorobyov}, E.~I., \& {Basu}, S. 2006, \apj, 650, 956

\bibitem[{{White} \& {Basri}(2003)}]{white03}
{White}, R.~J., \& {Basri}, G. 2003, \apj, 582, 1109

\bibitem[{{White} \& {Ghez}(2001)}]{white01}
{White}, R.~J., \& {Ghez}, A.~M. 2001, \apj, 556, 265

\bibitem[{{Whittet} {et~al.}(2004){Whittet}, {Shenoy}, {Clayton}, \&
  {Gordon}}]{whittet04}
{Whittet}, D.~C.~B., {Shenoy}, S.~S., {Clayton}, G.~C., \& {Gordon}, K.~D.
  2004, \apj, 602, 291

\bibitem[{{Zapatero Osorio} {et~al.}(2002){Zapatero Osorio}, {B{\'e}jar},
  {Pavlenko}, {Rebolo}, {Allende Prieto}, {Mart{\'{\i}}n}, \& {Garc{\'{\i}}a
  L{\'o}pez}}]{zapatero02}
{Zapatero Osorio}, M.~R., {B{\'e}jar}, V.~J.~S., {Pavlenko}, Y., {et~al.} 2002,
  \aap, 384, 937

\bibitem[{{Zhu} {et~al.}(2011){Zhu}, {Nelson}, {Hartmann}, {Espaillat}, \&
  {Calvet}}]{zhu11}
{Zhu}, Z., {Nelson}, R.~P., {Hartmann}, L., {Espaillat}, C., \& {Calvet}, N.
  2011, \apj, 729, 47

\end{thebibliography}

\clearpage

\begin{center}

\begin{deluxetable}{lccccc}
\tablewidth{0pt}
\tablecaption{Log of Observations
\label{tabobs}}
\tablehead{
\colhead{Object} &\colhead{RA}& \colhead{DEC}&\colhead{Telescope/ Instrument}& \colhead{Date of Obs}&\colhead{UT} \\
\colhead{}&\colhead{(J2000)}&\colhead{(J2000)}&&&\colhead{(hh:mm)}}
\startdata
SO 540&05 38 29.14& -02 16 15.7&Magellan/ MIKE&11-21-2009&07:07\\
&&&Magellan/ MagE&11-21-2009&06:09\\
&&&MDM/ 4k&11-22-2009&08:20\\
&&&SWIFT/ UVOT&03-11-2011\\
SO 1036&05 39 25.23& -02 38 22.0&Magellan/ MIKE&11-21-2009&07:28\\
&&&Magellan/ MagE&11-21-2009&06:25\\
&&&MDM/ 4k&11-22-2009&08:26\\
CVSO 58&05 29 23.26& -01 25 15.5&Magellan/ MIKE&11-22-2009&03:07\\
&&&Magellan/ MagE&11-22-2009&05:31\\
&&&MDM/ 4k&11-23-2009&05:44\\
CVSO 90&05 31 20.63& -00 49 19.8&Magellan/ MIKE&11-22-2009&06:47\\
&&&Magellan/ MagE&11-22-2009&06:34\\
&&&MDM/ 4k&11-22-2009&12:19\\
CVSO 107&05 32 25.79& -00 36 53.4&Magellan/ MIKE&11-22-2009&07:11\\
&&&Magellan/ MagE&11-22-2009&06:08\\
&&&MDM/ 4k&11-23-2009&05:50\\
CVSO 109&05 32 32.65& -01 13 46.1&Magellan/ MIKE&11-20-2009&05:56\\
&&&Magellan/ MagE&11-20-2009&03:27\\
&&&MDM/ 4k&11-21-2009&08:47\\
CVSO 206&05 24 41.03& 01 54 38.6&Magellan/ MIKE&11-22-2009&04:01\\
&&&Magellan/ MagE&11-22-2009&05:46\\
&&&MDM/ 4k&11-23-2009&05:38\\
OB1a 1630&05 26 55.37& 01 40 22.4&Magellan/ MIKE&11-22-2009&07:34\\
&&&Magellan/ MagE&11-22-2009&06:18\\
&&&MDM/ 4k&11-23-2009&05:44\\
&&&SWIFT/ UVOT&04-09-2010\\
\enddata
\end{deluxetable}

\begin{deluxetable}{lccccccccc}
\tablewidth{0pt}
\tablecaption{WTTS Templates
\label{tabwtts}}
\tablehead{
\colhead{Object}& \colhead{SpT*}&\colhead{$A_V$}&\colhead{Region}\\
}
\startdata
2MASS J05264681+0226039& M1&0.2&Orion OB1a\\
CHXR 48& M1.5& 1.1&Chamaeleon I\\
CVSO 127&M0.5& 0.4&Orion OB1b\\
CVSO 173&K7&0&Orion OB1b\\
SO 774& K7.5&1.6& $\sigma$ Ori\\
\enddata
\tablecomments{* Error on spectral type is $\pm$1 subclass}
\end{deluxetable}

\begin{deluxetable}{lcccccccccc}
\tablewidth{0pt}
\tablecaption{MDM and SWIFT Observed Magnitudes
\label{tabflux}}
\tablehead{
\colhead{Object}& \colhead{$U^{a}$}&\colhead{$V^{a}$}& \colhead{$R^a$}& \colhead{$I^a$}& \colhead{$UVW2^b$}& \colhead{$UVM2^b$}& \colhead{$UVW1^b$}& \colhead{$U^b$}& \colhead{$B^{b}$}& \colhead{$V^{b}$}} 
\startdata
SO 540     &16.03	&14.50	&13.64	&12.87	&17.58	&17.52	&16.57	&15.72	&15.66	&14.63	 \\
SO 1036   &15.58  	&14.72	&13.74	&12.77	&--		&--		&--		&--		&--		&--\\
CVSO 58  &15.55	&14.92	&14.16	&13.28 	&--		&--		&--		&--		&--		&--\\
CVSO 90  &15.16	&15.71	&14.85	&13.87	&--		&--		&--		&--		&--		&--\\
CVSO 107&16.04	&14.88	&13.98	&12.98	&--		&--		&--		&--		&--		&--\\
CVSO 109&15.67	&14.33 	&13.46	&12.45	&--		&--		&--		&--		&--		&--\\
CVSO 206&15.17	&14.39	&13.58	&12.76 	&--		&--		&--		&--		&--		&--\\
OB1a 1630&16.26	&15.60	&14.84	&13.94	&17.07	&16.56	&16.04	&15.91	&16.64	&16.03\\
\enddata
\tablecomments{$^a$photometry from MDM, $^b$photometry from SWIFT.}
\end{deluxetable}

\begin{deluxetable}{lccccccccc}
\tablewidth{0pt}
\tablecaption{CTTS Sample Properties
\label{tabprop}}
\tablehead{
\colhead{Object}& \colhead{SpT*}&\colhead{$r_V$}& \colhead{$r_I$}& \colhead{$A_V$}& \colhead{L}& \colhead{R}& \colhead{M}& \colhead{Region} \\
\colhead{}& \colhead{}&\colhead{}&\colhead{}&\colhead{(mag)}&\colhead{($\lsun$)}&\colhead{($\rsun$)}&\colhead{($\msun$)}& \colhead{}}
\startdata
SO 540     & K7		& 0.1$\pm$0.04	& 0.1$\pm$0.05	& 0.1$\pm$0.3		&0.9$\pm$0.1		&2.0$\pm$0.3		& 0.8$\pm$0.2	& $\sigma$ Ori$^1$\\
SO 1036   & M0.5	& 0.5$\pm$0.1		& 0.2$\pm$0.1		&0.5$\pm$0.5		&1.4$\pm$0.2		&2.9$\pm$0.5		& 0.5$\pm$0.1		& $\sigma$ Ori$^1$ \\
CVSO 58  & K7.5	& 0.8$\pm$0.1		& 0.4$\pm$0.07	& 0.8$\pm$0.5		& 0.5$\pm$0.1		&1.5$\pm$0.3		& 0.8$\pm$0.2	&  OB1b$^2$\\
CVSO 90  & M0.5	&1.5$\pm$0.3		&0.9$\pm$0.2		&0.0$\pm$0.4		&0.3$\pm$0.1		&1.5$\pm$0.2		&0.4$\pm$0.1		&OB1b$^2$\\
CVSO 107& K7.5	& 0.1$\pm$0.04	& 0.1$\pm$0.04	&0.7$\pm$0.4		&0.9$\pm$0.1		&2.0$\pm$0.3		&0.8$\pm$0.2		& OB1b$^2$\\
CVSO 109& K7.5	& 0.5$\pm$0.1		& 0.2$\pm$0.2		&0.8$\pm$0.5		& 1.5$\pm$0.2		&2.8$\pm$0.5		& 0.6$\pm$0.2	& OB1b$^2$\\
CVSO 206&  K7	& 0.6$\pm$0.2		& 0.3$\pm$0.1		&0.6$\pm$0.5		& 0.3$\pm$0.1		&1.1$\pm$0.2		& 0.8$\pm$0.2		& OB1a$^3$\\
OB1a 1630&M1.5	& 0.8$\pm$0.2		& 0.7$\pm$0.1		&0.0$\pm$0.2		& 0.1$\pm$0.1		&0.8$\pm$0.3		& 0.4$\pm$0.1	& OB1a$^4$\\
\enddata
\tablecomments{* Error on spectral type is $\pm$1 subclass. $^1$\citet{hernandez07b}, $^2$\citet{briceno05}, $^3$\citet{briceno07}, $^4$\citet{hernandez07a}}
\end{deluxetable}

\begin{deluxetable}{lcccccccc}
\tablewidth{0pt}
\tablecaption{Filling Factors for Multi-Component Model
\label{tabfill}}
\tablehead{
\colhead{Object} &\colhead{$f(10^{10})$}&\colhead{$f(3\times10^{10})$}&\colhead{$f(10^{11})$}&\colhead{$f(3\times10^{11})$}&\colhead{$f(10^{12})$}&\colhead{$f_{tot}$}&\colhead{$\mdot$ }&\colhead{$L_{acc}$ }\\
\colhead{}&\colhead{}&\colhead{}&\colhead{}&\colhead{}&\colhead{}&\colhead{ }&\colhead{($\msunyr$)}&\colhead{($\lsun$)}}
\startdata
SO 540	&	0	&0.04	&0	&0.0005	&0		&0.0405	&$8.5\times10^{-9}$&0.1\\
SO 1036	&	0.05	&0		&0	&0		&0.0002	&0.0502	&$2.0\times10^{-8}$&0.1\\
CVSO 58	&	0	& 0.03	&0	&0		&0.005	&0.035	&$1.6\times10^{-8}$&0.3\\		
CVSO 90 &	0.002&0		&0	&0.003	&0.001	&0.006	& $1.0\times10^{-8}$&0.08\\
CVSO 107&	0	&0		&0	&0		&0.0004	&0.0004	&$2.5\times10^{-9}$&0.03\\
CVSO 109&	0.1	&0		&0	&0		&0.0003	&0.1003	&$3.0\times10^{-8}$&0.2\\
CVSO 206&	0	&0		&0.03&0		&0.004	&0.034	&$7.4\times10^{-9}$&0.2\\
OB1a 1630&	0.005&0		&0	&0.002	&0.0003	&0.0073	&$7.7\times10^{-10}$&0.01\\
\enddata
\end{deluxetable}

\end{center}

\end{document}